\documentclass[12pt,epsf]{article}

\usepackage{graphicx}
\usepackage{amsmath,amssymb}
\usepackage{bm}
\usepackage{latexsym}
\allowdisplaybreaks%
\begin{document}
\baselineskip=0.7cm
\newcommand{\EQ}{\begin{equation}}
\newcommand{\EN}{\end{equation}}
\newcommand{\EQA}{\begin{eqnarray}}
\newcommand{\EQN}{\end{eqnarray}}
\newcommand{\EQAN}{\begin{eqnarray*}}
\newcommand{\EQNN}{\end{eqnarray*}}
\newcommand{\e}{{\rm e}}
\renewcommand{\theequation}{\arabic{section}.\arabic{equation}}
\newcommand{\Tr}{{\rm Tr}}
\newcommand{\lpartial}{\buildrel \leftarrow \over \partial}
\newcommand{\rpartial}{\buildrel \rightarrow \over 
\partial}
\newcommand{\np}{{\rm :}}
\newcommand{\llbracket}{[\hspace{-1.6pt}[}
\newcommand{\rrbracket}{]\hspace{-1.6pt}]}
\newcommand{\llbracketbar}{[\hspace{-1.5pt}[ \hspace{-7pt}\smallsetminus}
\newcommand{\rrbracketbar}{]\hspace{-1.5pt}]\hspace{-10pt}\smallsetminus}
\newcommand{\vertbar}{\hspace{3pt}|\hspace{-9pt}\smallsetminus\hspace{-3pt}}
\newcommand{\slashldelimit}{\hspace{-3pt}\smallsetminus}
\newcommand{\rhdllbra}{\triangleright\hspace{-8pt}[\hspace{-1.6pt}[ \hspace{4pt}}
\newcommand{\rhdrrbra}{\, \triangleright \hspace{-5pt}\rrbracket \,}
\renewcommand{\thesection}{\arabic{section}.}
\renewcommand{\thesubsection}{\arabic{section}.\arabic{subsection}}
\renewcommand{\figurename}{Fig.\@}
\makeatletter
\def\lesim{\mathrel{\mathpalette\gl@align<}}
\def\gtsim{\mathrel{\mathpalette\gl@align>}}
\def\gl@align#1#2{\lower.7ex\vbox{\baselineskip\z@skip\lineskip.2ex%
  \ialign{$\m@th#1\hfil##\hfil$\crcr#2\crcr\sim\crcr}}}
\makeatother
\makeatletter
\def\section{\@startsection{section}{1}{\z@}{-3.5ex plus -1ex minus 
 -.2ex}{2.3ex plus .2ex}{\large}} 
\def\subsection{\@startsection{subsection}{2}{\z@}{-3.25ex plus -1ex minus 
 -.2ex}{1.5ex plus .2ex}{\normalsize\it}}
\makeatother
\makeatletter
\def\lesim{\mathrel{\mathpalette\gl@align<}}
\def\gtsim{\mathrel{\mathpalette\gl@align>}}
\def\gl@align#1#2{\lower.7ex\vbox{\baselineskip\z@skip\lineskip.2ex%
  \ialign{$\m@th#1\hfil##\hfil$\crcr#2\crcr\sim\crcr}}}
\makeatother

@%
\def\thefootnote{\fnsymbol{footnote}}

\begin{flushright}
hep-th/yymmnnn\\
UT-KOMABA/07-6\\
May 2007
\end{flushright}
\vspace{0.3cm}

\begin{center}
{\Large  Field Theory of Yang-Mills Quantum Mechanics for 
D Particles}

\vspace{0.4cm}

Tamiaki {\sc  Yoneya}
\footnote{
E-mail address: tam\_at\_hep1.c.u-tokyo.ac.jp
}

\vspace{0.3cm}

{\it Institute of Physics, University of Tokyo \\
Komaba, Meguro-ku, Tokyo 153-8902, Japan}

\vspace{1cm}
Abstract
\end{center}

We propose a new field-theoretic framework for formulating the 
non-relativistic quantum mechanics of D particles (D0 branes) in a Fock space of 
U($N$) Yang-Mills theories with all different $N$ simultaneously. 
D-particle field operators, which create and annihilate 
a D particle and 
hence change the value of $N$ by one, are 
defined. The base space of these D particle fields is a (complex) 
vector space of infinite dimension. The gauge invariance of 
Yang-Mills quantum mechanics is reinterpreted 
as a quantum-statistical symmetry, which is taken into account 
by setting up 
a novel algebraic and projective structure in the formalism. 
Ordinary physical observables of Yang-Mills theory, obeying 
the standard algebra, 
are expressed as bilinear forms of the D-particle fields. 
Together with the open-closed string duality, our new formulation 
suggests a trinity of three different but dual 
viewpoints of string theory.

\newpage
\section{Introduction}

The idea of a quantized field is a precise expression of 
the primordial duality between waves and particles in quantum theory. 
Historically,  it stemmed  from the quantization of the 
classical electromagnetic field on the side of waves, 
while on the particle side 
 from the second quantization of the configuration-space 
formulation of non-relativistic particle quantum mechanics. 
As a unification of two classically different concepts, the notion  
of a quantized field should be 
taken as more fundamental than its ancestors. 
It is therefore natural that string field theory \cite{sft,witten,hikko, zwiebach} has 
been developed from the early days of string theory 
towards its non-perturbative formulation. 
Indeed, even though  during the first 
two decades of their initial development 
the various versions of string field theory represent mere rewritings  
of the world-sheet CFT dynamics of strings in terms 
of ordinary Feynman rules, the situation has been
 changing in recent years: 
String field theory, especially the version 
of open-string field theories first proposed in the seminal  
work Ref.\cite{witten},  
 is now playing an important role 
  in the studies of some crucial aspects  of non-perturbative 
string physics, such as tachyon condensation, albeit still at 
a classical level \cite{recent}.

With the advent of D-branes, 
it became recognized that the natures of open and closed 
string field theories are quite different: Open strings, and 
hence open-string fields, 
should be understood as the collective degrees of freedom for 
describing D-branes, while closed strings 
are the bulk degrees of freedom corresponding to 
gravitons and their partners for which D-branes can 
be regarded as sources or as soliton-like objects. 
In closed string field theory, they can emerge 
as classical solutions (with or without sources) for the equations of motion. In principle, 
these two descriptions must be connected 
through the duality between open and closed strings, and 
constitute a foundation for gauge-gravity correspondence. 
However, there has been only limited success in formulating  
the open-closed string duality from this viewpoint. 
For example, it has not been clarified how to extract the 
bulk degrees of freedom of closed strings concretely from 
open-string field theory and vice versa, 
although some suggestive 
properties have been discovered, such as those 
related to lump solutions, 
to be interpreted as D-branes 
in the proposal of the so-called {\it `vacuum'} string field theory 
\cite{vsft} in bosonic string theory. 

In ordinary local quantum field theory, it is well known that 
there is an analogous 
phenomenon , 
 which provides us a clear and rigorous example of a 
duality related to solitons. That is the duality 
\cite{mduality} 
between the massive Thirring model and the sine-Gordon model 
in two-dimensional spacetime. In this case, the kink (and 
anti-kink) solitons  
of the latter are described  by a massive Dirac field whose 
interaction is described by the non-linear 4-fermi term 
in the former. These two 
descriptions can be translated into each other in both 
directions through bosonization or fermionization. 
It should also be mentioned that in two dimensions the massless 
QED, the Schwinger model \cite{schwinger}, provides 
a suggestive analog to the open-closed string duality, 
in the sense that the one-loop quantum effect of the 
massless Dirac field 
gives a pole singularity corresponding to a composite massive 
scalar field. This is reminiscent of the phenomenon 
that the one-loop amplitudes of open strings give rise to 
pole singularities associated with the propagation of a closed string. 

In these examples of the formulations of duality, 
it is crucial that the soliton degrees of 
freedom are treated as 
 a quantized field, namely as 
 the Dirac field for which
the bosonization gives the sine-Gordon field 
(or the free massive scalar field 
in the case of the Schwinger model). 
The Dirac field corresponds to D-branes 
described collectively by open strings, while the 
elementary bosonic fields, as the analog of a closed string field, 
are obtained from bilinear forms of the Dirac field.

These considerations provide  motivation for 
attempting to construct a certain field-theory-like formalism for treating 
D-branes, in the hope of clarifying the fundamental open-closed 
string duality and thereby seeking 
new languages for non-perturbative 
string/M physics.  It seems worthwhile to explore 
the possibility of 
field operators creating and annihilating 
D-branes, within the setup of the Fock space of D-branes. 
Although some of the mathematical ideas usually employed 
in describing D-branes, 
such as the K-theory approach, 
involve certain aspects of  
changing the number of D-branes 
(and anti-D-branes) from the viewpoint 
of their topological properties, 
they do not seem from a physics viewpoint 
to be concrete enough as a formalism 
of treating real dynamics. 
The purpose of the present work is to initiate a more 
physical approach, introducing D-brane 
field operators explicitly. We hope to take a first step towards  
our goal by treating the case of D0-branes, D-particles, 
within the low-energy approximation such that they are 
properly described by U($N$) (super) Yang-Mills quantum mechanics \cite{ymqm}. 
Note that the open-string field theory with Chan-Paton 
symmetry and Yang-Mills quantum 
mechanics as its low-energy effective theory are 
both first-quantized formulations for D particles, 
though they are second-quantized theories 
from the viewpoint of open strings. 
Thus, we  introduce the Fock space of the Yang-Mills 
theory in $0+1$ dimensions by treating all the different $N$
 in a completely unified way. We can think of 
  many desirable extensions, for example, 
 including anti-D-branes and covariantizating the formalism 
 both in 10 and 11 dimensions \cite{bfss} in the context of M-theory. 
 Such attempts would perhaps require the 
 inclusion of all massive degrees of freedom of open strings. 
This is left hopefully for future works. 
 
 In the literature of D-brane 
 Yang-Mills theories, 
it has been claimed 
that the Yang-Mills quantum mechanics 
is {\it already} second-quantized. However, in precise terms,  
this is not true, since Yang-Mills quantum mechanics 
is yet a configuration-space formulation of D particles, 
in which the basic degrees of freedom are matrix variables 
whose diagonal components represent nothing but the 
coordinates of D particles. 
Also, by a {\it `second'} quantization of Yang-Mills theory, one might 
imagine to introduce just a collection of symmetrized tensor products of 
an arbitrary number of Hilbert spaces of the
 same Yang-Mills theory. However, this is {\it not} what we 
want, as will become evident through
 our discussion given in the next section.  

In a previous work \cite{I}, the present author discussed 
a similar question in the case of D3-branes within 
the restriction of  the 1/2-BPS condition. Unfortunately, however, 
under the latter condition, we could only deal with the 
two-point correlation functions (and the so-called extremal 
correlators) of a special class of operators 
which are protected from interactions. 
In the present work, we do not require such additional restrictions,  
other than the low-energy approximation and the exclusion of 
anti-D particles. Thus the situation 
is quite different from that of the previous work. Yet there is an  important lesson which we shall follow in the present work. 
We expect an entirely new mathematical structure 
for treating D-brane fields. The reason for this is, as we discuss in detail 
in the following sections, that the symmetry of the configuration-space 
formulation underlying multi-D-brane states  
is not the permutation symmetry of usual multi-particle 
quantum mechanics.  Rather, it must be replaced by the
 U($N$) gauge symmetry of Yang-Mills theory.  
In other words, D-branes can be treated neither as ordinary bosons 
nor fermions, and their fields do not have any classical counterparts. 
From the viewpoint of ordinary quantum field theory, this is 
a drastic leap, which forces us to go beyond 
the standard framework of quantum field theory. 
What we explore in the 
following is mainly the problem of how to deal with such a quantum system 
with a {\it continuous} statistical symmetry. 
As we argue, this requires that we invent a novel   
structure for the operator algebra of D-particle fields, 
which to the author's knowledge has never been conceived. 

This paper is  organized as follows. 
In section 2, we discuss the problems related to 
the  continuous statistical symmetry and 
explain our approach for separating the matrix degrees of 
freedom such that they are suitable for the idea of a Fock space 
of Yang-Mills theory with all different $N$. We then 
introduce D-particle field operators and explain 
their novel properties, which are clearly beyond 
the ordinary canonical framework of quantum field theories.  
In section 3, we show how to construct various physical 
observables corresponding to the gauge invariant operators 
of Yang-Mills theory as 
bilinear forms of the D-particle fields.  As many of 
our concepts and notation must be entirely alien to 
the reader, some elementary details of the calculation are 
presented in the text for the purpose of 
avoiding confusion. In a first reading, the reader 
may skip these details of the calculations. 
The extension of the present formalism to the case of super 
Yang-Mills theory is discussed briefly in section 4. 
Finally, section 5 is devoted to concluding remarks. 

\section{The Fock space of D-particle quantum mechanics}

\subsection{Preliminaries}
In trying to formulate a second-quantized field theory of 
D-branes whose fundamental degrees of freedom in the first-quantized 
formulation 
are matrix variables, one of the puzzling questions is how to 
deal with gauge symmetry. We need a unified treatment for 
all sizes of matrices.  A typical case among those with which we 
are already familiar is the $c=1$ matrix model 
with a single Hermitian $N\times N$ matrix variable $X_{ab}(t)$, 
restricted to its singlet sector. 
In this case, we can first diagonalize the matrix variable and treat 
the $N$ eigenvalues $\lambda_a$ as independent degrees of freedom, whereby the residual gauge symmetry is reduced to the 
permutation symmetry of the eigenvalues. Then, 
by taking into account the gauge-volume factor, 
the Vandelmonde determinant, 
originating in the integration measure for the matrix variable, the model 
can be treated as a system of $N$ free fermions whose 
coordinates are those eigenvalues. 
Thus, in this particular example, the problem of 
second-quantizing matrix 
systems with all different $N$  is reduced to the standard one. 
It is shown in \cite{I}
that this mechanism can be extended to all the SO($6$)  
multiplets  of 1/2-BPS operators of ${\cal N}=4$ SYM$_4$ if we associate 
the SO($6$) degrees of freedom with the so-called Cuntz algebra. 

Unfortunately, this procedure cannot be extended to 
situations with two or more matrix degrees of freedom without 
some strong restrictions,  such as the 1/2-BPS condition. 
In the case of D0-branes, D particles,  in $d$-dimensional space, we 
have to deal with $d$ matrix coordinates on their world lines, 
$
X^i_{ab}(t)
$ ( $ i\in (1, 2, \ldots, d) ;
a, b, \ldots \in (1, 2, \ldots, N)$  )
and their super-partners, 
where $N$ is the 
number of D0-branes or RR-charge and $i$ represents the  
spatial directions. In particular, 
the positions of D-particles 
are described by the diagonal elements, 
whereas the off-diagonal elements represent 
the lowest modes of open strings connecting 
different D-branes. 
The theory must respect the Chan-Paton gauge symmetry 
$
X^i\rightarrow UX^{i}U^{-1}, \, \,  U\in \mbox{U($N$)}
$. 
Throughout this paper, the spatial dimensions should be 
understood as $d=9$, but we use 
notation such as $d^d x$  for the integration measure for 
bosonic variables, 
since the formalism is valid for any dimensions, until 
we take into account the supersymmetry in section 4. 

One conceivable extension of the above procedure 
to multi-matrix cases 
would be to treat the entire set 
of gauge invariants, replacing the role of 
eigenvalues,  as collective field variables \cite{collective}. 
There have been numerous attempts of 
 reformulating various multi-matrix models and 
 Yang-Mills theories in terms of Wilson-loop-like operators. 
 However, in such approaches, there is an important difficulty 
 that is often ignored: It is difficult 
 to separate the independent dynamical degrees of freedom from 
 the set of all possible invariants. The number of
 such independent degrees of 
 freedom, of course, depends on the rank $N$ of the gauge group.  
 This difficulty\footnote{
 In an early work
 \cite{yoneold}
  of the present author, an approach for 
 circumventing this difficulty was studied with the goal of developing 
 a field theory of Wilson loops. The reader is  referred to that work and  papers cited therein 
 for some early attempts related to this subject. 
 } may be ignored if we are interested only 
 in the usual large $N$ limit. But it makes 
 the precise treatment of the Fock space of matrix models with 
 variable $N$  almost hopeless. 
 
 In this section, we propose a new approach which is entirely 
 different from any previous attempts. Let us begin 
  by briefly recalling the ordinary second quantization 
 in the elementary non-relativistic quantum mechanics of identical particles.  Here, 
 the $N$-body wave function at a fixed time defined on 
  the configuration space 
${\cal H}_N$  is 
replaced by a state vector $|\Psi \rangle$  in the Fock space $
{\cal F}=\sum_N \bigoplus {\cal H}_N$ as 
\[
\Psi(x_1, x_2, \ldots, x_N)
\]
\begin{equation}
\Rightarrow |\Psi\rangle 
=\Big(
\prod_{i=1}^N \int d^dx_i\Big) 
\Psi(x_1, x_2, \ldots, x_N)
\psi^{\dagger}(x_N)\psi^{\dagger}(x_{N-1}) \cdots \psi^{\dagger}(x_1)|0\rangle
\end{equation}
where the field operators define mappings between the configuration 
spaces ${\cal H}_N$ with different $N$: 
\[
\psi^{\dagger}(x): {\cal H}_N \rightarrow {\cal H}_{N+1}
, \]
\[
\psi(x): {\cal H}_N\rightarrow {\cal H}_{N-1}.\]
The quantum-statistical symmetry is expressed as 
\begin{equation}
\psi^{\dagger}
(x_{N})\psi^{\dagger}(x_{N-1}) \cdots \psi^{\dagger}(x_1)|0\rangle
=\psi^{\dagger}(x_{P(N)})\psi^{\dagger}(x_{P(N-1)}) \cdots \psi^{\dagger}(x_{P(1)})|0\rangle
\label{usualstatistics} 
\end{equation}
and 
\begin{equation}
\psi(y)\psi^{\dagger}(x_N)\cdots \psi^{\dagger}(x_1)|0\rangle
={1\over (N-1)!}\sum_P \delta^d(y-x_{P(N)})
\psi^{\dagger}(x_{P(N-1)})\ldots 
\ldots \psi^{\dagger}(x_{P(1)})|0\rangle
\label{usualannihilation}
\end{equation}
where the summation is over all permutations  $P:(12\ldots N)\rightarrow 
(i_1i_2\ldots i_N), P(k)=i_{k}$. Here, 
we have assumed bosons for definiteness. In supersymmetric theory, 
fermions can be taken into account within the same 
 formalism by considering superfields introducing 
super-coordintes appropriately. In what follows, we sometimes
 suppress the spatial indices. 

Of course, the entire structure is reformulated 
as a representation theory 
of the canonical commutation relations of the field 
operators acting on the Fock vacuum $|0\rangle, 
\psi(x)|0\rangle =0$:  
\[
[\psi(x), \psi^{\dagger}(y)]=\delta^d(x-y), \quad 
[\psi(x), \psi(y)]=0=[\psi^{\dagger}(x), 
\psi^{\dagger}(y)].
\]
 For the purposes of the present paper, however,  it is important to note that the relations
 \eqref{usualstatistics} and \eqref{usualannihilation} are actually 
 sufficient for expressing various observables, such as 
 the number operator,  $\int d^dx \, \psi^{\dagger}(x)\psi(x)$,  of multi-particle systems 
 in terms of bilinear forms of operator fields, $\psi(x)$ and $ \psi^{\dagger}(x)$. 
 We attempt to extend the definitions \eqref{usualstatistics} 
 and \eqref{usualannihilation} directly, but {\it not} the canonical algebra,  at least in the present work.

 In the case of $N$ D-particles, the configuration space 
is replaced as \[
(x_1, x_2. \ldots, x_N) \rightarrow \{X_{ab}=\bar{X}_{ab}\}
\]
and the permutation symmetry is replaced by the gauge symmetry under 
\[
X_{ab}\rightarrow 
(UXU^{-1})_{ab}. 
\]
Comparing with the case of ordinary particles, we recognize two unusual features: 
\begin{enumerate}
\item[{\it 1}.] The increase in the number of the degrees of freedom of 
the mapping ${\cal H}_N
\rightarrow {\cal H}_{N+1}$ is
$d_N\equiv 
d(2N+1)=d((N+1)^2-N^2)$, instead of $d=d(N+1)-dN$, which is 
independent of $N$. 
\item[{\it 2}.] The statistical symmetry is a continuous group, the 
adjoint representation of  
U($N$), instead of the discrete group of permutations $\{P\}$.
\end{enumerate}
Evidently,  the second quantization of D-particles 
 cannot be performed within the standard framework.

Our proposal for dealing with this situation is as follows. 
The feature {\it 1} indicates that  D-particle fields $\phi^{\pm}[z,\bar{z}]$,
creating or annihilating 
 a D-particle,  must be 
defined on a base space with an infinite number of 
coordinate components, since $d_N\rightarrow \infty$ as 
$N\rightarrow \infty$. But, if they act on a state with a 
definite number $N$ of D-particles, only the finite 
number, $d_N$, of them must be activated, and 
the remaining ones should be treated as dummy variables.  
In terms of the matrix coordinates, 
we first redefine components of these infinite dimensional 
space as 
\begin{equation}
z^{(b)}_a=X_{ab} =\bar{X}_{ba}  \quad \mbox{for}  \quad b\ge a, 
\end{equation}
which is to be interpreted as the $a$-th component 
of the (complex) coordinates of the $b$-th D-particle.  
The  assumption here is that the field algebra and its 
representation should be set up such that we can effectively ignore 
the components 
$z^{(b)}_a$ and $\bar{z}^{(b)}_a$ with $a>b$ for the 
$b$-th operation in adding 
D-particles. 
 Hence, the matrix variables are embedded into the sets of  arrays  
of infinite-dimensional complex vectors $(z_1=x_1+iy_1,z_2=x_2+iy_2, \ldots)$.   Note that the upper indices with braces discriminate 
the D-particles by ordering them, 
whereas the lower indices without braces represent 
the components of the infinite-dimensional 
coordinate vector $(z,\bar{z})=\{z_1,\bar{z}_1, z_2, \bar{z}_2, \ldots \}$ for each D-particle. 

In  the example of the 4-body case,  we have  $4\times 4$ matrices 
of the form
\begin{equation}
X=\begin{pmatrix} 
x_1^{(1)} & z_1^{(2)}& z^{(3)}_1 & z^{(4)}_1 \cr 
\bar{z}_1^{(2)} & x_2^{(2)} & z^{(3)}_2 & z^{(4)}_2 \cr
\bar{z}_1^{(3)} & \bar{z}_2^{(3)} & x_3^{(3)} & z^{(4)}_3 \cr
\bar{z}_1^{(4)} & \bar{z}_2^{(4)} & \bar{z}^{(4)}_3 & x_4^{(4)} \cr
\end{pmatrix}.
\end{equation}
Then, in terms of the infinite-dimensional vectors, this is rearranged as 
a set of four complex vectors, 
\[
z^{(1)}=
\begin{pmatrix}
x_1^{(1)}\cr \cdot \cr  \cdot \cr \cdot  \cr
 \cdot \cr 
\cdot \cr \cdot \cr
\end{pmatrix},\quad 
z^{(2)}=
\begin{pmatrix}
z_1^{(2)}\cr x_2^{(2)}\cr \cdot \cr \cdot \cr 
 \cdot \cr 
\cdot \cr \cdot \cr
\end{pmatrix}, \quad 
z^{(3)}=
\begin{pmatrix}
z_1^{(3)}\cr z_2^{(3)}\cr x_3^{(3)}\cr \cdot \cr
 \cdot \cr 
\cdot \cr \cdot \cr
\end{pmatrix}, \quad 
z^{(4)}=
\begin{pmatrix}
z_1^{(4)}\cr z_2^{(4)}\cr z_3^{(4)}\cr x_4^{(4)}\cr \cdot \cr 
\cdot \cr \cdot \cr
\end{pmatrix}
\]
together with their complex conjugates. The dots indicate inifinitely 
many dummy components. Of course, both $X$ and $(z, \bar{z})$ are 
spatial vectors, with the corresponding indices being suppressed.

Thus the D-particle fields creating and annihilating 
a D particle must be defined conceptually as
\[
\phi^+ : \, |0\rangle \rightarrow \phi^+[z^{(1)}, \overline{z}^{(1)}]|0\rangle 
\, \rightarrow \phi^+[z^{(2)}, \overline{z}^{(2)}]\phi^+[z^{(1)}, 
\overline{z}^{(1)}]|0\rangle \, \rightarrow \cdots ,
\]
\[\hspace{0.9cm}
\phi^- : \, 
0 
\, \leftarrow|0\rangle \, \leftarrow
\phi^+[z^{(1)}, \overline{z}^{(1)}]|0\rangle \leftarrow 
\phi^+[z^{(2)}, \overline{z}^{(2)}]\phi^+[z^{(1)}, 
\overline{z}^{(1)}]|0\rangle \leftarrow \cdots .
\]
The manner in which the degrees of freedom are 
added (or subtracted) is illustrated in Fig. 1. 
\begin{figure}[htbp]
\begin{center}
\includegraphics[width=15cm]{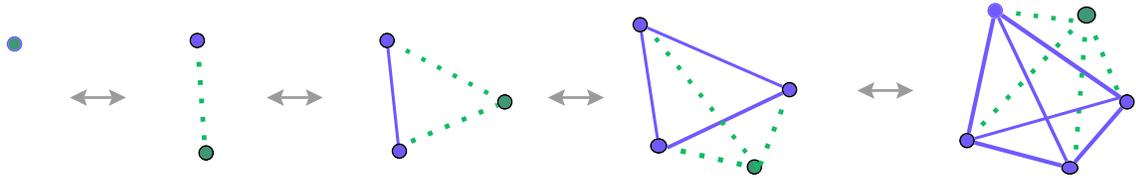}
\caption{\footnotesize{
The D-particle coordinates and the open strings mediating them are 
denoted by blobs and lines connecting them, respectively. 
The real lines are open-string degrees of freedom 
which have been created before the 
latest operation of the creation field operator, while the dotted lines 
indicate those created by the last operation. The arrows 
indicate the operation of 
creation (from left to right) and annihilation 
(from right to left) of D-particles.
}}
\end{center}
\vspace{-0.3cm}
\end{figure}
As we explain below, the multiplication rules of the 
D-particle field operators are actually {\it not} associative, 
nor commutative, and hence we need some special 
notation for expressing $N$-body states with $N\ge 3$. 
But for the moment, let us avoid such complications 
for the purpose of explaining only the 
 key ideas in the case of  simple 1-body and 2-body states. 

\subsection{Projection condition and the creation operator}
The conditions on the effective coordinate degrees of freedom 
are assumed  to be the following recursive requirements:
\[
\partial_{y_1^{(1)}}\phi^+[z^{(1)}, \bar{z}^{(1)}]|0\rangle =0, 
\quad 
\partial_{z^{(1)}_k}\phi^+[z^{(1)}, \bar{z}^{(1)}]|0\rangle =
\partial_{\bar{z}^{(1)}_k}\phi^+[z^{(1)}, \bar{z}^{(1)}]|0\rangle 
=0 \quad 
\mbox{for}\, \quad 
k\ge 2, 
\]
\[
\partial_{y_2^{(2)}}\phi^+[z^{(2)}, \bar{z}^{(2)}]\phi^+[z^{(1)}, 
\bar{z}^{(1)}]
|0\rangle =0, 
\quad 
\partial_{z^{(2)}_k}\phi^+[z^{(2)}, \bar{z}^{(2)}]
 \phi^+[z^{(1)}, \overline{z}^{(1)}]|0\rangle 
\]
\begin{equation}
=
\partial_{\bar{z}^{(2)}_k}\phi^+[z^{(2)}, \bar{z}^{(2)}]
 \phi^+[z^{(1)}, \overline{z}^{(1)}]|0\rangle =0 \quad 
\mbox{for}\, \quad 
k\ge 3, \quad \mbox{etc.}
\label{reduction}
\end{equation}
They require that the non-vanishing results of the 
creation operator sending a $N$-body 
state to a $N+1$-body state have nontrivial coordinate dependence 
 only for finite-dimensional sub-manifolds, as we have defined 
 in the previous subsection, 
and have only zero-mode (in the sense of the conjugate 
momentum space) dependencies 
on other components $y_N^{(N)}$ and 
$z^{(N)}_{k}$ with $k\ge N+1$. 

To express these conditions concisely for the general 
case, it is convenient to introduce new notation. 
Let $P_1$ denote the operation of projection which 
reduces a generic complex vector 
$(w_1, w_2, w_3, \dots )$ to $(r_1, 0, 0, \ldots)$ 
with $w_n=r_n+ i s_n$, 
\begin{equation}
P_1:  \, \,  (w_1, w_2, w_3, \ldots ) 
\rightarrow(r_1, 0, 0, \ldots), 
\end{equation}
and more generally as
\begin{equation}
P_n : \, \, (w_1, w_2, \ldots, w_{n-1}, w_n, w_{n+1}, \ldots) 
\rightarrow (w_1, w_2, \ldots, w_{n-1}, r_n, 0, 0, 0, \ldots),
\label{momentumprojection}
\end{equation}
satisfying 
$
P_nP_m=P_mP_n=P_m $ for $n\ge m. $
Equivalently, we can express the latter conditions as 
\begin{equation}
Q_nQ_m= \delta_{nm}Q_n, 
\end{equation}
defining
$
Q_n=P_{n}-P_{n-1} $ for $n\ge 2$, 
and $Q_1=P_1.
$
We use this notation with the following convention: 
For an arbitrary function $f[w,\bar{w}]$ of complex vectors, 
the projected functions are defined either as 
\begin{equation}
w_k f[P_n, \overline{P_n w}]=0 \quad \mbox{for}\quad k>n, \quad 
s_n f[P_n w, \overline{P_n w}]=0 
\end{equation}
or as
\begin{equation}
\partial_{w_k} f[\tilde{P}_n, \overline{\tilde{P}_n w}]=0 \quad \mbox{for}\quad k>n, \quad 
\partial_{s_n} f[\tilde{P}_n w, \overline{\tilde{P}_n w}]=0. 
\end{equation}
For brevity, we have denoted the results of projection with respect to
 the conjugate momentum $p_n$ as defined 
in \eqref{momentumprojection} 
by $\tilde{P}_nz$. When we consider the projection 
in the coordinate representation, we simply write  it as  
$P_nz$ without the tilde. This is also the case for our annihilation operator 
$\phi^-$, to be defined later.  It should be 
kept in mind that 
for {\it non}-dummy components, there is clearly no distinction 
between projection operators with or without the tilde in the base space. 

The conditions \eqref{reduction} for the creation field acting 
sequentially upon the vacuum can be expressed by introducing 
an infinite set of projection operators ${\cal P}_n$ $(n=1, 2, \ldots)$ 
in the D-particle Fock space 
satisfying 
\begin{equation}
\phi^+[z, \bar{z}]{\cal P}_n=
{\cal P}_{n+1}\phi^+[\tilde{P}_nz, \tilde{P}_n\bar{z}] . 
\label{projectioncreation}
\end{equation}
Unlike the projections $P_n$ and $\tilde{P}_n$ in the base 
space,  the Fock-space projectors ${\cal P}_n$ with different 
$n$ are assumed to be orthogonal:
\begin{equation}
{\cal P}_n{\cal P}_m=\delta_{mn}{\cal P}_n. 
\end{equation}
This is consistent with the orthogonality of the Fock space states with 
different D-particle numbers. 
The action of the projection 
operator ${\cal P}_n$ on the creation D-field $\phi^+$ makes the 
projection in the base space of the arguments 
of the field operators with respect to its momentum space, 
as manifest in \eqref{reduction}. 
The vacuum is assumed to be invariant under the 
projection ${\cal P}_1$ but  annihilated by  ${\cal P}_n$ with 
$n\ge 2$:
\begin{equation}
{\cal P}_1|0\rangle =|0\rangle, \quad {\cal P}_2|0\rangle=
{\cal P}_3|0\rangle=\cdots = 0 . 
\label{vacuumprojection}
\end{equation}
Then, the projection conditions \eqref{reduction} are succinctly 
summarized as 
\begin{equation}
\phi^+[z^{(2)}, \bar{z}^{(2)}]\phi^+[z^{(1)}, \bar{z}^{(1)}]|0\rangle 
={\cal P}_{3}\phi^+[\tilde{P}_2z^{(2)}, \tilde{P}_2\bar{z}^{(2)}]
\phi^+[\tilde{P}_1z^{(1)}, \tilde{P}_1\bar{z}^{(1)}]|0\rangle .
\end{equation}
Extension to general $N$-body states is obvious. When 
we introduce the dual Fock space below, the dual vacuum $\langle 0|$ 
is assumed to satisfy the same conditions  
as \eqref{vacuumprojection}, 
\begin{equation}
\langle 0|{\cal P}_1 =\langle 0|, \quad \langle 0|{\cal P}_2=
\langle 0|{\cal P}_3=\cdots = 0 . 
\label{dualvacuumprojection}
\end{equation}

These projection 
conditions should be regarded as the quantum analog of the projector 
property realized by 
the  lump solutions which are believed to represent 
D-branes in the vacuum string field theory. As is well known, 
such a feature has previously been observed in the case of 
non-commutative solitons \cite{noncommutative}. In these cases, 
the projector property is exhibited by the {\it  classical} 
solutions themselves. Such classical 
solutions must be related to the collective coordinates 
for solitons, which should constitute the base space for soliton-field operators in a full-fledged quantum theory. 

Because of these conditions, it is often convenient and 
most economical to 
write these multi-D-particle states 
by indicating only the components of the coordinates on 
which the field operators have nontrivial dependencies 
and suppressing the dummy components, as 
\[
{\cal P}_2\phi^+[X_{11}]|0\rangle , \quad 
{\cal P}_3\phi^+[X_{12}, X_{21}, X_{22}]\phi^+[X_{11}]|0\rangle , \quad \mbox{etc.}
\]

\subsection{Gauge-statistics conditions and the annihilation operator}

We can now impose conditions to 
account for the second new feature of our D-particle Fock  space. 
The feature {\it 2} of the D-particle Fock space 
mentioned in the subsection {\it 2.1},  is trivial for a 1-body state,  
being effective only for states with $N\ge 2$. We demand 
for $N=2$ that 
\begin{equation}
{\cal P}_3\phi^+[X_{12}, X_{21}, X_{22}]\phi^+[X_{11}]|0\rangle 
={\cal P}_3\phi^+[X^{U(2)}_{12}, X^{U(2)}_{21}, 
X^{U(2)}_{22}]\phi^+[X^{U(2)}_{11}]|0\rangle, 
\label{gaugestatistics2}
\end{equation}
with obvious extension for $N\ge 3$, where 
\begin{equation}
X^{U(N)}\equiv U(N)XU(N)^{-1}.
\label{adjointtransform}
\end{equation}
Here $U(N)$ is an arbitrary unitary $N\times N$ 
matrix. 
This realizes a natural generalization of the condition 
\eqref{usualstatistics}  to the present {\it continuous} 
statistical symmetry through the replacement of the 
permutation symmetry $\{P\}$ by the continuous group U($N$). 
Although we are forced to abandon the simple commutativity 
of the creation operators corresponding to the usual 
Bose-Einstein statistics, we impose instead an 
extremely strong gauge-symmetry constraint. 
 Compared to ordinary particles, D particles are not only 
indistinguishable from each other in a  stringent 
sense, but also they are mutually permeable, or, {\it osmotic}. 
We call this new quantum statistics  `permeable statistics'.  

If we use the notation of the projection operators introduced 
above, we can express the {\it non}-dummy components 
of complex vector coordinates by 
making the  replacement 
\begin{equation}
P_nz^{(n)}\rightarrow P_nX Q_n, 
\quad 
P_n\bar{z}^{(n)}\rightarrow Q_nX P_n, 
\end{equation}
using an abstract matrix notation 
for the right-hand side, which should be understood as
\begin{equation}
(k| P_nXQ_n|\ell)
=\delta_{\ell n}X_{k\ell} \quad \mbox{for} \quad k\le n, 
\end{equation}
\begin{equation}
(\ell| Q_nX P_n|k)
=\delta_{\ell n}X_{\ell k} \quad \mbox{for} \quad k\le n, 
\end{equation}
or as 
\begin{equation}
Q_n|k)=\delta_{nk}|k), \quad (k|Q_n=
\delta_{nk}( k|. 
\end{equation}
Here, the variable $X$ is a generic infinite-dimensional operator 
whose projection is assumed to be the corresponding finite-dimensional 
matrix variable. In terms of this notation, 
the transformation \eqref{adjointtransform} is nothing but 
the change of basis 
\begin{equation}
|n) \rightarrow |n)'=\sum_{m=1}^N U(N)^{-1}_{mn}|m)
\quad ( n| \rightarrow (n|'=\sum_{m=1}^N U(N)_{nm}(m|
\end{equation}
for $n \le N$, under which $P_N$ is invariant
\begin{equation}
P_N=\sum_{n=1}^N|n) (n|=\sum_{n=1}^N
|n)' ( n|'. 
\end{equation}
Thus, the gauge-statistics condition 
requires that the $N$-body states be 
invariant under the change of the maximal set 
$\{|1), |2), \ldots, |N)\}$ of the 
basis for  the $N\times N$ matrices of the $N$-body states.

Our next task is to define the  annihilation operator $\phi^-[
z, \bar{z}]$ for D particles. Its action must be consistent with 
the above properties of the creation operator. 
First, for the vacuum, we have
\begin{equation}
\phi^-[z,\overline{z}]|0\rangle =0. 
\end{equation}
Then, for 1-body state, we define
\begin{equation}
\phi^-[z, \overline{z}]\phi^+[z^{(1)}, \overline{z}^{(1)}]|0\rangle 
=\delta(x_1-x^{(1)}_1)\delta(y_1)\prod_{k\ge 2}\delta^2 (z_k)|0\rangle, 
\end{equation}
where $\delta^2(z) =\delta(z)\delta(\bar{z})$. 
We use the following abbreviation for the product of  an 
infinite number of $\delta$-functions:
\begin{equation}
\sigma_\ell\equiv \delta(y_{\ell})\prod_{k>\ell}\delta^2(z_k). 
\end{equation}
 Note that here and in what 
follows, we suppress  
the dimension parameter $d$ on the $\delta$-functions 
for spatial vectors. Thus, we write  $\delta(y) =\delta^d(y), \delta^2(z)=
\delta^{2d}(z)$, etc. 

The attentive reader might wonder why we here 
use the position space $\delta$-functions for the dummy 
components of $(z, \bar{z})$, in spite of the fact that 
the projection of the dummy components for the creation fields is 
done in the {\it momentum} space. The reason is that 
for the annihilation field, we assume the projection to take  the  following form
\begin{equation}
{\cal P}_n\phi^-[P_nz, P_n\bar{z}]=\phi^-[z, \bar{z}]{\cal P}_{n+1}, 
\label{projectionannihilation}
\end{equation}
where the projections $P_nz$ with respect to the basis space of 
the complex vectors are effective 
in the {\it coordinate} representation, in contrast to the creation field. 
Such mutually conjugate projections of dummy components 
for the creation and annihilation operators are 
necessary in order to ensure the reduction of the number of 
degrees of freedom in a well-defined way 
for $N$-body D-particle states from infinite dimensions. By so doing, 
the parts of the Hamiltonian and other observables
 become null for the dummy variables 
as  desired. Otherwise we would encounter various annoying 
infinities in making the reductions. We do not claim 
uniqueness for our construction: For example, 
 in our present formulation 
in flat space, the theory is invariant under  
  $(z^i_n, \bar{z}_n^i) \rightarrow (z^i_n, \bar{z}_n^i)
+(c^i,c^i)$, as we show below.  Thus we can at least 
shift the origin of the infinite-dimensional vector space 
accordingly. 

Beyond the above remarks on the projection condition, 
the two properties of the annihilation operator 
stated above  need no further explanation. 
A  non-trivial feature
 appears from 2-body states. As a natural extension 
of \eqref{usualannihilation}, we define the action of the 
annihilation field to the 2-body state by
\[
\phi^-[z,\overline{z}]\phi^+[z^{(2)}, \bar{z}^{(2)}]
\phi^+[z^{(1)}, \bar{z}^{(1)}]|0\rangle 
=\phi^-[z,\overline{z}]{\cal P}_3
\phi^+[X_{12}, X_{21}, X_{22}]\phi^+[X_{11}]|0\rangle
\]
\begin{equation}
={\cal P}_2
\int [dU(2)] \delta(x_2-
X^{U(2)}_{22})\delta^2(z_1-X^{U(2)}_{12})
\sigma_2
\phi^+[X^{U(2)}_{11}]|0\rangle.
\end{equation}
Here, the group 
integration measure $[dU(N)]$ is normalized as 
\begin{equation}
\int [dU(N)] =N. 
\end{equation}
An important characteristic of this definition is that 
we introduced the integration $[dU(N)]$  over the $U(N)$ $(N=2)$ 
transformations explicitly in order to preserve the gauge-statistics 
condition of the original state, $\phi^+[z^{(2)}, \bar{z}^{(2)}]
\phi^+[z^{(1)}, \bar{z}^{(1)}]|0\rangle 
$, before acting with the annihilation field. 
This, of course, corresponds to the summation $\sum_P$ 
over different permutations $\{P\}$ in \eqref{usualannihilation} 
of the $N$-body states of ordinary particles, and it 
replaces the role of the canonical commutator 
$[\psi(x), \psi^{\dagger}(y)]=\delta(x-y)$ 
between annihilation and creation fields for ordinary particles. 

\subsection{Non-associativity}
Before going on to formulate the properties 
for D-particle fields for the general $N$-body state, it is appropriate here to 
give further analysis concerning 
the nature of the decomposition of matrix 
variables into sets of infinite-dimensional complex vectors. 
Due to the existence of the off-diagonal components 
of matrix variables, we cannot, generically,  reduce the 
action of a subgroup of a gauge transformation group 
to a smaller subset of variables of matrix components. 
In contrast to this, in the case of the 
permutation group, any 
permutation within a subset of particles does not 
affect the remaining particle degrees of freedom. 
This implies that, when we consider a mapping from 
$N$-body  to $(N+1)$-body states of D particles, the gauge-statistics condition 
of the former cannot be embedded into the latter 
without affecting the new degrees of freedom 
added by the operation of creating a D-particle. 

For example, suppose that a gauge transformed 
$N\times N$ matrix $X^{U(N)}=U(N)XU(N)^{-1}$ of an 
$N$-body state is embedded naturally 
in an $(N+1)\times (N+1)$ matrix of an $(N+1)$-body state as
\[
X^{\hat{U}(N)}\equiv \begin{pmatrix}
 X^{U(N)}_{1,1}  & X^{U(N)}_{1,2}  &  \cdots & X^{U(N)}_{1,N} &X_{1, N+1}\cr
 X^{U(N)}_{2,1} & X^{U(N)}_{2,2} & \cdots & X^{U(N)}_{2,N} & X_{2, N+1}\cr
 \cdot & \cdot & \cdot & \cdot &\cdot \cr
 \cdot & \cdot & \cdot & \cdot &\cdot \cr
 \cdot & \cdot & \cdot & \cdot &\cdot \cr
 X^{U(N)}_{N,1} & X^{U(N)}_{N, 2} & \cdots & X^{U(N)}_{N, N} 
 &X_{N, N+1}\cr
 X_{N+1,1}& X_{N+1, 2} & \cdots & X_{N+1, N} & X_{N+1, N+1}
\end{pmatrix}.
\]
But this is different from  the $(N+1)\times (N+1)$ matrix 
\[
X^{U(N+1;N)}, 
\]
which is obtained with the $(N+1)\times (N+1)$ 
unitary transformation containing the $N\times N$ transformation 
$U(N)$ as a block-diagonal matrix 
as 
\begin{equation}
U(N+1;N)\equiv  \begin{pmatrix}
U(N)& 0\cr
          0 & 1 \cr
\end{pmatrix} .
\label{embedding}
\end{equation}
In the latter case, the $N$-th column and row must be 
replaced, respectively,  by $(a, b \le N)$
\begin{equation}
z^{(N+1)}_a=X_{a, N+1}\rightarrow U(N)_{ab}X_{b,N+1}, 
\quad \bar{z}^{(N+1)}_a=X_{N+1, a}\rightarrow X_{N+1, b}U(N)^{-1}_{ba} , 
\label{replacement}
\end{equation}
compared with $X^{\hat{U}(N)}$ in the above expression. 
Thus, a U($N$) adjoint transformation of  the $N\times N$ sub-matrix 
must necessarily be associated with a U($N$) transformation 
of the added column and row of the $(N+1)\times (N+1)$ matrix 
in which the original $N\times N$ matrix is embedded. 

This means that an $N$-ple product of the D-particle creation fields 
cannot be decomposed into a form which is 
composed of products with smaller numbers of them for  states 
with smaller $N$. 
Thus, the algebra of D-particle creation operators are, so to speak, 
 {\it maximally} non-associative.  
In order to express this situation explicitly, we denote 
the $N$-body state created by the product of $N$ creation fields 
in the form
\begin{equation}
|N[X]\rangle \equiv {\cal P}_N\llbracket\phi^+[X_N]\phi^+[X_{N-1}]\cdots 
\phi^+[X_1]\rrbracket|0\rangle,
\label{nbodystate}
\end{equation}
employing a special delimiter symbol $\llbracket *** \rrbracket$ 
for the ordered products ($***$)  of field operators. 
By definition, the state \eqref{nbodystate} satisfies the gauge-statistics condition
\begin{equation}
\llbracket\phi^+[X_N]\phi^+[X_{N-1}]\cdots 
\phi^+[X_1]\rrbracket|0\rangle
=
\llbracket\phi^+[X^{U(N)}_N]\phi^+[X^{U(N)}_{N-1}]\cdots 
\phi^+[X^{U(N)}_1]\rrbracket|0\rangle .
\end{equation}
Note that here (and mostly in what follows) we 
use abbreviated notation for 
matrix variables as the arguments for the field operators: 
The complex vectors for which each creation field 
has nontrivial dependence are denoted by 
$(z^{(n)}, \bar{z}^{(n)}) \rightarrow X_n$, 
with $n=1, 2, \ldots, N$ for the $N$-body state.  
Thus, for example, $X_N$ collectively represents the set of $2N+1$ 
elements given by
 \[
(X_{1,N}, X_{N,1}, X_{2, N}, X_{N, 2}, \ldots, 
X_{N-1,N}, X_{N, N-1}, X_{N,N}).\]

At this juncture, we add one side remark concerning the decomposability 
of the multiplication rule for creation operators. Consider 
a U($N$) matrix of the form
\[
U(N;n)\equiv 
\begin{pmatrix}
1_{N-n} & 0 \cr
0 & U(n) \cr
\end{pmatrix}, 
\]
with $U(n)$ being an $n\times n$ unitary matrix. 
The adjoint action of this matrix on an $N$-body matrix variables
 by this matrix affects 
its columns and rows corresponding only to 
the $n$-th, $(n+1)$-th, $\ldots$, $N$-th creation fields, {\it i.e.}
$\phi^+[X_n], \ldots, \phi^+[X_N]$. This suggests 
the
 possibility of nesting the non-associative $N$-ple multiplication of the 
creation operators 
 into a particular set of successive multiplications  from 
 the left to the right as 
\[
\llbracket\phi^+[X_N]\phi^+[X_{N-1}]\cdots 
\phi^+[X_1]\rrbracket|0\rangle
\sim 
((\cdots ((\phi^+[X_N]\phi^+[X_{N-1}])\phi^+[X_{N-2}]) \cdots )
\phi^+[X_1])|0\rangle. 
\] 
This is not suitable for studying recursive multiplications 
of field operators from right to left, corresponding to the 
creation of D-particles one by one, yielding states with increasing 
number particle number. 
But it is certainly 
suggestive of a certain algebraic characterization 
of the D-particle field operators. We leave further investigation 
of this and related aspects of  D-particle fields 
for future works.

\subsection{Two classes of operations in the Fock space}
With the above feature in mind, 
the operation 
$
|N[X]\rangle \xrightarrow{\phi^+[z]}|(N+1)[X]\rangle 
$
that sends the $N$-body state 
$|N[X]\rangle$ to the $N+1$-body state by 
inserting a new creation field $\phi^+[X_{N+1}]$ into the 
bracket symbol $\llbracket *** \rrbracket$, 
\begin{equation}
|(N+1)[X]\rangle
={\cal P}_{N+2}\llbracket\phi^+[X_{N+1}]\phi^+[X_N]\phi^+[X_{N-1}]\cdots 
\phi^+[X_1]
\rrbracket |0\rangle, 
\end{equation}
is represented by the new symbol 
$
\triangleright\llbracket \, \, =\triangleright
\hspace{-6pt}\llbracket \hspace{3pt}
$
as 
\[
\phi^+[z](|N[X]\rangle) 
\equiv \phi^+[z] 
\triangleright
|N[X]\rangle 
\]
\[
={\cal P}_{N+2}
\phi^+[X_{N+1}] 
\triangleright\hspace{-9pt}\llbracket \hspace{4pt}
 \phi^+[X_N]\phi^+[X_{N-1}]\cdots 
\phi^+[X_1] \rrbracket|0\rangle
\]
\begin{equation}
={\cal P}_{N+2}\llbracket\phi^+[X_{N+1}]\phi^+[X_N]\phi^+[X_{N-1}]\cdots 
\phi^+[X_1]\rrbracket |0\rangle.
\end{equation}
Thus, this new symbol 
represents the operation of adding a new blob and off-diagonals,  
depicted by the dotted lines in Fig. 1. 
If this process is formally interpreted as the  multiplication of 
 the bracket $\llbracket *** \rrbracket$ by creation fields from 
 the left, it is 
{\it not} invariant under the $U(N)$ transformation in the 
sense of the original $N$-body state, 
\[
\phi^+[X_{N+1}]\rhdllbra 
\phi^+[X^{U(N)}_N]\phi^+[X^{U(N)}_{N-1}]\cdots 
\phi^+[X^{U(N)}_1] \rrbracket |0\rangle
\]
\[
\ne 
\phi^+[X_{N+1}]\rhdllbra 
\phi^+[X_N]\phi^+[X_{N-1}]\cdots 
\phi^+[X_1] \rrbracket |0\rangle, 
\]
but instead 
satisfies the $N+1$-body gauge-statistics condition, of which 
a special case is 
\[
|(N+1)[X^{\hat{U}(N)}]\rangle ={\cal P}_{N+2}\phi^+[X_{N+1}]\rhdllbra 
\phi^+[X^{U(N)}_N]\phi^+[X^{U(N)}_{N-1}]\cdots 
\phi^+[X^{U(N)}_1]\rrbracket |0\rangle
\]
\begin{equation}
={\cal P}_{N+2}
\phi^+[X^{U(N+1;N)}_{N+1}]
\rhdllbra 
\phi^+[X_N]\phi^+[X_{N-1}]\cdots 
\phi^+[X_1] \rrbracket |0\rangle
\label{hatinvariance}
\end{equation}
\[
={\cal P}_{N+2}
\llbracket \phi^+[X^{U(N+1;N)}_{N+1}] \phi^+[X_N]\phi^+[X_{N-1}]\cdots 
\phi^+[X_1]\rrbracket |0\rangle, 
\]
where $X^{U(N+1;N)}_{N+1}$ indicates the replacement 
given in \eqref{replacement}. 

This peculiarity does not create any immediate inconsistency 
{\it if} we forbid this operation from appearing
 in {\it physical} processes, 
even though  care must be taken  in treating these operators. 
This signifies that our D-particle fields themselves 
should not be regarded directly as physical observables. 
There is a similar situation in ordinary field theory; recall that 
 any non-singlet fields in general gauge field theories 
cannot be regarded directly as gauge-invariant observables, 
but are still useful as intermediate tools for constructing observables. 
 As a matter of fact, the creation of a D particle 
not accompanied by the simultaneous creation of an anti-D-particle 
is forbidden, due to the RR-charge conservation from the 
bulk viewpoint also. 

It is convenient to classify various operations in our D-particle Fock space 
according to the following criterion. 
If an operation ${\cal O}$ acting on an $N$-body state 
preserves the gauge-statistics condition, {\it i.e.}
\begin{equation}
{\cal O}|N[X]\rangle ={\cal O}|N[X^{U(N)}]\rangle, 
\label{uninvariance}
\end{equation} 
the operation ${\cal O}$ is defined to be A-class. Otherwise, 
it will be called B-class. Thus the D-particle creation 
field acting to the right is B-class. 
However, it is possible that 
a B-class operation can become A-class when it is 
combined with other operations. In general, any operator representing 
observables must be A-class, though the converse need not be 
true, depending on further criteria for `physical' operations. 
We  see below that integrated 
local bilinear forms of the creation and annihilation fields 
are A-class in general, and they obey the ordinary algebra 
of gauge-invariants when acting upon states. 
 
Let us now  fix the definition of the annihilation operator by 
extending that given above for the 2-body case 
to general $N$-body states. 
Using the new symbols, the action of the annihilation field 
for a general $N$-body state is expressed as 
\[
\phi^-[z,\overline{z}]|N[X]\rangle 
=\phi^-[z,\overline{z}]
{\cal P}_{N+1}\llbracket \phi^+[X_N]
\phi^+[X_{N-1}]\cdots
\phi^+[X_1]\rrbracket
|0\rangle 
\]
\[=\phi^-[z, \bar{z}]{\cal P}_{N+1}\phi^+[X_N]
\rhdllbra 
\phi^+[X_{N-1}]\phi^+[X_{N-2}]\cdots 
\phi^+[X_1] \rrbracket |0\rangle
\]
\[=
\int [dV(N)]\, \delta(x_N-
X^{V(N)}_{NN})\delta^2(z_1-X^{V(N)}_{1N})
\delta^2(z_2-X^{V(N)}_{2N})
\cdots \delta^2(z_{N-1}-X^{V(N)}_{N-1, N})
\sigma_N
\]
\begin{equation}
\times \, {\cal P}_N
\llbracket \phi^+[X^{V(N)}_{N-1}]\phi^+[X^{V(N)}_{N-2}]
\cdots \phi^+[X_{1}^{V(N)}]\rrbracket
|0\rangle . 
\label{Nannihilation3}
\end{equation}
Note that the delimiter in the first line is still 
``$\llbracket$", corresponding to the fact 
that the operation of annihilation applied to ket-states preserves the 
gauge statistics condition of the $N$-body states, owing to the 
U($N$)-integration, and hence it is A-class when acting to the right. 

\subsection{Dual states}
Using a general gauge invariant wave function $\Psi[X]=\Psi[X^{U(N)}]$ in the configuration space,  
a general $N$-body state vector in the Fock space 
takes the form
\begin{equation}
|\Psi\rangle = \int \frac{[d^{d}X]}{[dU(N)]} \Psi[X]|N[X]\rangle,
\label{generalstate}
\end{equation}
where  in the integration over the matrix coordinates, the 
gauge redundancy is completely removed by the denominator 
representing the gauge volume, $[dU(N)]$. 
To complete the formalism, it is necessary to define dual states 
that are conjugate to 
these ket-states. The bra-vacuum satisfies 
\[
\langle 0| \phi^+[z, \bar{z}]=0, \quad \langle 0|0\rangle =1,
\]
with the projection condition \eqref{dualvacuumprojection}. 
Then the  general dual $N$-body states are given by 
\begin{equation}
\langle \Psi|
=
\int [d^{2d}\hat{z}]_{N}
\int\frac{[d^{d}Y]}{[dU[N]]}\overline{\Psi[Y]}
\langle N[Y;\hat{z}]|{\cal P}_N
\label{dualstate}
\end{equation}
in terms of  the dual $N$-body basis state
\begin{equation}
\langle N[Y;\hat{z}]|=\langle 0|\llbracket
\phi^-[Y_1;\hat{z}^{(1)}]\phi^-[Y_2;
\hat{z}^{(2)} ]\cdots 
\phi^-[Y_N;\hat{z}^{(N)}]\rrbracket, 
\end{equation}
where we have denoted the components with nontrivial 
dependencies by $Y_n$. The symbols $\hat{z}^{(n)}$ collectively 
represent the 
remaining dummy components of the complex vector 
coordinates
\[
\hat{z}^{n} : \, \{y^{(n)}_n; z^{(n)}_{n+1},z^{(n)}_{n+2}, \ldots, ;
 \bar{z}^{(n)}_{n+1}, \bar{z}^{(n)}_{n+2},\ldots\} , 
\]
with respect to which the state vector has only $\delta$-function supports at zero, by the definition 
of the action of the annihilation fields, owing to the 
projection condition \eqref{projectionannihilation}. 
Correspondingly, the symbol $[d^{2d}\hat{z}]_{N}$ in 
\eqref{dualstate} indicates 
integrations over them. However,  it should be recalled that the 
wave function $\overline{\Psi[X]}$ is simply the 
complex conjugate of the Yang-Mills wave function $\Psi[X]$,  
which is independent of the dummy components. 
Therefore, when the dual bra-states  are coupled with the ket-states, 
the integration $\int [d^{2d}\hat{z}]_N$ over 
these remaining components simply gives 1. 
Hence, we can in practice suppress these variables 
for notational brevity, as we do in what follows,  
unless otherwise specified. 

Using a similar convention similar to that used in
 the case of the ket-states, we define the new right delimiter 
 symbol 
 $
 \rrbracket\triangleright =\rhdrrbra  
 $
 and denote the operation to the left sending 
the dual $(N-1)$-body basis vector to the $N$-body basis vector as 
\begin{equation}
\langle N[Y;\hat{z}]|=
\langle 0|\llbracket\phi^-[Y_1;\hat{z}^{(1)}]\phi^-[Y_2;\hat{z}^{(2)}]\cdots 
\phi^-[Y_{N-1};\hat{z}^{(N-1)}]
\rhdrrbra 
\phi^-[Y_N;\hat{z}^{(N)}] , 
\end{equation}
or,  by suppressing the dummy variables, simply as 
\[
\langle N[Y]|=
\langle 0|\llbracket\phi^-[Y_1]\phi^-[Y_2]\cdots 
\phi^-[Y_{N-1}]\rhdrrbra 
\phi^-[Y_N]
\]
\[=
\langle 0|\llbracket\phi^-[Y_1]\phi^-[Y_2]\cdots 
\phi^-[Y_{N-1}]
\phi^-[Y_N]\rrbracket
\]
using the same convention as for the ket-states. 
The scalar product of two states can be computed 
using our rules. It is evident that they are non-vanishing only 
for states with the same numbers of D particles. 
For 1-body states, we obviously have 
\[
\langle \Psi'|\Psi\rangle =
\int d^dY_{11}\int d^dX_{11}
\overline{\Psi'[Y]}\Psi[X]\delta(Y_{11}-X_{11}). 
\]
For general $N$-body 
states, we have
\begin{equation}
\langle \Psi'|\Psi\rangle =N!\int \frac{[d^{d}Y]}{[dU(N)]}
\int \frac{[d^{d}X]}{[dU(N)]}\overline{\Psi'[Y]}\Psi[X]
\delta [Y-X]. 
\label{internalproduct}
\end{equation}
 The $\delta$-function 
for the matrices $\delta[Y-X]$ is an $dN^2$-dimensional 
$\delta$-function with respect to the $dN^2$ integration 
measure for the matrix coordinates, equating all the $dN^2$ real components 
of two $d$-dimensional $N\times N$  hermitian matrices $X$ and $Y$. 

Indeed, after integration over the dummy 
vector components, which simply gives  1, we have  
\[
\int\frac{[d^dY(N)]}{[dU(N)]}\int\frac{[d^dX(N)]}{[dU(N)]}
\, \overline{\Psi'[Y]}
\Psi[X]
\langle N[Y]|N[X]\rangle =
\]
\[
\int\frac{[d^dY(N)]}{[dU(N)]}\int\frac{[d^dX(N)]}{[dU(N)]}
\, \overline{\Psi'[Y]}
\Psi[X]
\]
\[
\times 
\langle 0|\llbracket\phi^-[Y_1]\phi^-[Y_2]\cdots 
\phi^-[Y_{N-1}]\rhdrrbra\phi^-[Y_N]\phi^+[X_N]
\rhdllbra
\phi^+[X_{N-1}]\phi^+[X_{N-2}]\cdots 
\phi^+[X_1]\rrbracket |0\rangle
\]
\[
=
\int\frac{[d^dY(N)]}{[dU(N)]}\int\frac{[d^dX(N)]}{[dU(N)]}
\, \overline{\Psi'[Y]}
\Psi[X]
\int [dV(N)]
\langle (N-1)[Y]|(N-1)[X_N^{V(N)}]\rangle 
\delta[Y_N-X_N^{V(N)}]
\]
\[
=
\int\frac{[d^dY(N)]}{[dU(N)]}\int\frac{[d^dX(N)]}{[dU(N)]}
\, \overline{\Psi'[Y]}
\Psi[X]
\int [dV(N)]
\langle (N-1)[Y]|(N-1)[X]\rangle 
\delta[Y_N-X_N]
\]
\[
=N
\int\frac{[d^dY(N)]}{[dU(N)]}\int\frac{[d^dX(N)]}{[dU(N)]}
\, \overline{\Psi'[Y]}
\Psi[X]
\langle (N-1)[Y]|(N-1)[X]\rangle 
\delta[Y_N-X_N].
\]
Note that here the $\delta$-function $\delta[Y_N-X_N]$ 
is one of $d(2(N-1)+1)$ dimensions.  
This leads to the above conclusion, 
\eqref{internalproduct}, utilizing the fact that the 
wave functions $\Psi[X]$ and $\Psi'[Y]$ and 
the states $| n[X]\rangle$ and  $\langle n[Y] |$ $(n=1, 2, \ldots, N)$ 
are all {\it independently} invariant under gauge transformations. 
It is also to be noted that the $\delta$-functions $\delta[Y_n-X_n]$ 
can be replaced by 
$\delta[Y_m^{U(N, n)}-X_m^{U(N, n)}]$ $(m\ge n)$ 
in the case that we  embed 
an arbitrary U($n$) matrix into U($N$), as in 
the definition \eqref{embedding} with $n=N-1$ and its obvious 
generalization to general $n \, (<N)$. This is necessary to reduce 
the last line to lower body states recursively. 
 We have thus seen that 
our Fock space admits the ordinary probability interpretation, 
even though we have to deal with the non-associative 
rule for the multiplication of field operators. 

The above property of the dual states also allows us to assume the action 
of the creation operator to bra-states from the right to the left 
in the following form :
\[
\langle N[Y]|\phi^+[z,\bar{z}]=\langle 0|\llbracket\phi^-[Y_1]\phi^-[Y_2]\cdots 
\phi^-[Y_{N-1}]
\phi^-[Y_N]\rrbracket
\phi^+[z, \bar{z}]{\cal P}_{N-1}
\]
\[=
\langle 0|\llbracket\phi^-[Y_1]\phi^-[Y_2]\cdots 
\phi^-[Y_{N-1}]\rhdrrbra 
\phi^-[Y_N]\phi^+[z,\bar{z}]{\cal P}_{N-1}
\]
\[
=
\int [dV(N)]\delta(x_N-Y^{V(N)}_{NN})
\delta^2(z_1-Y_{1N}^{V(N)})\delta^2(z_2-Y_{2N}^{V(N)})
\cdots \delta^2(z_{N-1}-Y_{N-1, N}^{V(N)})
\]
\begin{equation}
\times \langle 0|\llbracket\phi^-[Y^{V(N)}_1]
\phi^-[Y^{V(N)}_2]\cdots 
\phi^-[Y^{V(N)}_{N-1}]\rrbracket.
\end{equation}
It is to be noted that this expression is 
independent of $y_N$ and of $z_n$ and $\bar{z}_n$ with $n>N$,  
in accordance with the projection condition 
\eqref{projectioncreation}
satisfied by the creation operator $\phi^+[z, \bar{z}]$. 

These results show that the classification of 
operators defined in the previous subsection
 actually depends on the 
directions of the operations: The creation operator 
is B-class when acting from the left to the right, but  A-class 
when acting from the right to the left. Conversely, the annihilation operator is A-class from the left to the right and B-class from the right to the left. In this way, the apparent asymmetry between the creation and annihilation operators is resolved by taking into account 
the dual Fock space. 
Operators corresponding to physical processes {\it must} 
be A-class in both directions.

\section{Bilinears as physical operators}
\setcounter{equation}{0}
We are now ready to discuss how to express ordinary gauge-invariant 
physical observables including the Hamiltonian in terms of our 
D-particle field operators. As in the ordinary second quantization, 
they are essentially given by bilinear forms of D-particle field 
operators, 
\begin{equation}
\langle \phi^+, F\phi^-\rangle 
\equiv 
\int [d^{2d}z]\, \phi^+[z, \bar{z}]\triangleright  F\phi^-[z, \bar{z}] .
\end{equation}
We include the symbol $\triangleright$ here  to explicitly  indicate that we are using the rule of multiplication 
of the creation operator 
$\phi^+$ to the right and of the annihilation operator 
$\phi^-$ to the left, simultaneously, following 
the discussion in the previous section.  
The symbol $F$ can in general represent operators with respect to the vector coordinates $(z, \bar{z})$. 
Obviously, these bilinear forms 
commutes with the projection conditions, {\it i.e.}
\begin{equation}
[{\cal P}_n , \langle \phi^+, F\phi^-\rangle]=0, 
\end{equation}
and also are A-class in both directions of their 
operation. Since their properties when acting on states are 
essentially 
symmetric with respect to ket-states and bra-states, it is 
sufficient to study their actions on ket-states. 

\subsection{Bilinear operators without derivatives}
Let us first consider the case $F=F[z, \bar{z}]$,  which does not involve 
differential operators in $z$ and $\bar{z}$. 
By studying the action of this bilinear form on a general $N$-body ket-state $|N[X]\rangle$, we find 
\[
\int [d^{2d}z]\, \phi^+[z, \bar{z}]\triangleright  F\phi^-[z, \bar{z}]
|N[X]\rangle
\]
\[
=
\int [d^{2d}z]F[z, \bar{z}]\int [dU(N)]\, \delta(x_N-
X^{U(N)}_{NN})\delta^2(z_1-X^{U(N)}_{1N})
\delta^2(z_2-X^{U(N)}_{2N})
\cdots \delta^2(z_{N-1}-X^{U(N)}_{N-1, N})\sigma_N
\]
\[
\times \, \phi^+[z, \bar{z}]
{\cal P}_{N-1}\rhdllbra \phi^+[X^{U(N)}_{1, N-1}, X^{U(N)}_{2, N-1}, \ldots, 
\bar{X}^{U(N)}_{1,N-1}, \bar{X}^{U(N)}_{2, N-1}, 
\ldots, X_{N-1, N-1}^{U(N)}]
\cdots \phi^+[X_{11}^{U(N)}]\rrbracket|0\rangle 
\]
\[
={\cal P}_{N}\int [dU(N)]
\, 
F[X_{1, N}^{U(N)}, 
X_{2, N}^{U(N)}, \ldots, \bar{X}_{1,N}^{U(N)}, 
\bar{X}_{2,N}^{U(N)}, \ldots, X_{N,N}^{U(N)}, 
0, \ldots]
\]
\[
\times \llbracket \phi^+[X_{1, N}^{U(N)}, 
X_{2, N}^{U(N)}, \ldots, \bar{X}_{1,N}^{U(N)}, 
\bar{X}_{2,N}^{U(N)}, \ldots, X_{N,N}^{U(N)}]
\]
\[
\times 
\phi^+[X^{U(N)}_{1, N-1}, X^{U(N)}_{2, N-1}, \ldots, 
\bar{X}^{U(N)}_{1,N-1}, \bar{X}^{U(N)}_{2, N-1}, 
\ldots, X_{N-1, N-1}^{U(N)}]
\cdots \phi^+[X_{11}^{U(N)}]\rrbracket |0\rangle 
\]
\begin{equation}
=
\int [dU(N)]\, F[X_{1, N}^{U(N)}, 
X_{2, N}^{U(N)}, \ldots, \bar{X}_{1,N}^{U(N)}, 
\bar{X}_{2,N}^{U(N)}, \ldots, X_{N,N}^{U(N)}, 
0, \ldots]|N[X]\rangle,
\end{equation}
where in the last equality we have used the gauge-statistical 
symmetry of the general $N$-body state $|N[X]\rangle 
=|N[X^{U(N)}]\rangle$. 

\vspace{0.2cm}
\noindent
{\it Examples}:

\noindent
The simplest example is 
the number operator with $F=1$, 
\begin{equation}
\langle \phi^+, \phi^-\rangle |N[X]\rangle = N |N[X]\rangle. 
\end{equation}
The next simplest case, 
$
F_n=z^i_n , 
$
gives 
\begin{equation}
\langle \phi^+, z_n \phi^-\rangle |N[X]\rangle 
=\int [dU(N)](U(N) X^iU(N)^{-1})_{n, N}|N[X]\rangle = 
\Tr(X^i)\delta_{n, N} |N[X]\rangle, 
\end{equation}
using the formula 
\begin{equation}
\int [dU(N)] U(N)_{ab}U(N)^{-1}_{cd}=\delta_{ad}\delta_{bc}. 
\end{equation}
Note that for $n>N$, the projection condition for the 
annihilation operator is responsible for the vanishing result, 
while for $n<N$, the group integration is. 
Thus, we  have
\begin{equation}
\langle \phi^+,\sum_{n=1}^{\infty} z_n \phi^-\rangle |N[X]\rangle 
=\Tr(X^i) |N[X]\rangle .
\label{1storder}
\end{equation}
Actually, $F=\bar{z}_n^i$ gives the same expression,
\begin{equation}
\langle \phi^+,\sum_{n=1}^{\infty} \bar{z}_n \phi^-\rangle |N[X]\rangle 
=\Tr(X^i) |N[X]\rangle .
\label{1storderbar}
\end{equation}

For a quadratic form, 
\[
F=\sum_{n=1}^{\infty} z^i_n \bar{z}_n^j\equiv \bar{z}^j\cdot z^i
\]
which is manifestly invariant under the infinite-dimensional 
global transformation $z_n \rightarrow (Uz)_n$, with 
$ U$ being an arbitrary unitary U($\infty$) matrix,  
we arrive at the integral
\[
\int [dU(N)]
(U(N)X^jU(N)^{-1})_{N, n}(U(N) X^iU(N)^{-1})_{n, N}
\]
\[
=\int[dU(N)](U(N)X^jX^iU(N)^{-1})_{N,N}
=\Tr(X^jX^i),
\]
giving
\begin{equation}
\langle \phi^+, \bar{z}^j\cdot z^i\phi^-\rangle |N[X]\rangle 
=\Tr(X^jX^i)|N[X]\rangle.
\label{2ndorder}
\end{equation}

An example of 3rd-order invariants is obtained by considering 
$
F_n=(\bar{z}^i\cdot z^j)z^k_n.
$
Acting with the corresponding bilinear form  
on an $N$-body basis state, we have the $U(N)$ 
integral 
\[
\int [dU(N)] U(N)_{Na}(X^iX^j)_{ab}U(N)_{bN}^{-1}
U(N)_{nc}X_{cd}^kU(N)^{-1}_{dN}
\]
\[
=\int [dU(N)]U(N)_{Na}U(N)_{nc}U(N)_{bN}^{-1}U(N)^{-1}_{dN}
(X^iX^j)_{ab}X_{cd}^k
\]
\[
=
\Big(\frac{N}{N^2-1}(\delta_{NN}\delta_{nN}\delta_{ab}\delta_{cd}
+\delta_{NN}\delta_{nN}\delta_{ad}\delta_{cb})
-\frac{1}{N^2-1}(\delta_{NN}\delta_{nN}\delta_{ab}\delta_{cd}
+\delta_{NN}\delta_{nN}\delta_{ad}\delta_{cb})\Big)
\]
\[
\times 
(X^iX^j)_{ab}X_{cd}^k
\]
\[
=
\delta_{nN}\frac{1}{N+1}\Big(
\Tr(X^iX^j)\Tr(X^k)+\Tr(X^iX^jX^k)
\Big)
\]
which gives
\begin{equation}
\delta_{nN}\frac{1}{N+1}\Tr([X^i,X^j]X^k)|N[X]\rangle 
=\langle \phi^+, (\bar{z}^i\cdot z^j-\bar{z}^j\cdot z^i)z^k_n\phi^-\rangle
|N[X]\rangle. 
\end{equation}
Here use has been made of the integration formula
\[
\int [dU(N)] \, U(N)_{a_1b_1}U(N)_{a_2b_2}
U(N)^{-1}_{c_1d_1}U(N)^{-1}_{c_2d_2}
\]
\[
=
{N\over N^2-1}(\delta_{a_1d_1}\delta_{a_2d_2}\delta_{b_1c_1}
\delta_{b_2c_2}+ \delta_{a_1d_2}\delta_{a_2d_1}\delta_{b_1c_2}
\delta_{b_2c_1})
\]
\begin{equation}
-{1\over N^2-1}(
\delta_{a_1d_2}\delta_{a_2d_1}\delta_{b_1c_1}
\delta_{b_2c_2}
+\delta_{a_1d_1}\delta_{a_2d_2}\delta_{b_1c_2}
\delta_{b_2c_1}). 
\end{equation}
Thus, for arbitrary $N$, we can use the expression  
\begin{equation}
\sum_{n=1}^{\infty}\langle \phi^+, (\bar{z}^i\cdot z^j-\bar{z}^j\cdot z^i)z_n^k\phi^-\rangle |N[X]\rangle 
=
\frac{1}{N+1}\Tr([X^i,X^j]X^k)|N[X]\rangle 
\end{equation}
similarly to the case of the 1st-order invariant \eqref{1storder}. 
This expression is suitable when we consider the so-called 
Myers term.  

As is evident from the derivation, the last
 expression for odd invariants is not 
unique. For instance, we can instead consider 
$
F_n=(\bar{z}^i\cdot z^j)\bar{z}_n^k. 
$
Then, we have the U($N$) integral
\[
\int [dU(N)] U(N)_{Na}(X^iX^j)_{ab}U(N)_{bN}^{-1}
U(N)_{Nc}X_{cd}^kU(N)^{-1}_{dn}
\]
\[
=\int [dU(N)]U(N)_{Na}U(N)_{Nc}U(N)_{bN}^{-1}U(N)^{-1}_{dn}
(X^iX^j)_{ab}X_{cd}^k
\]
\[
=
\Big(\frac{N}{N^2-1}(\delta_{NN}\delta_{nN}\delta_{ab}\delta_{cd}
+\delta_{NN}\delta_{nN}\delta_{ad}\delta_{cb})
-\frac{1}{N^2-1}(\delta_{NN}\delta_{nN}\delta_{ab}\delta_{cd}
+\delta_{NN}\delta_{nN}\delta_{ad}\delta_{cb})\Big)
\]
\[
\times 
(X^iX^j)_{ab}X_{cd}^k
\]
\[
=
\delta_{nN}\frac{1}{N+1}\Big(
\Tr(X^iX^j)\Tr(X^k)+\Tr(X^iX^jX^k)
\Big).
\]
Thus we have an identity for two odd-degree bilinear operators  
as the action on states. 
\begin{equation}
\langle \phi^+, (\bar{z}^i\cdot z^j)\bar{z}_n^k\phi^-\rangle
=\langle \phi^+, (\bar{z}^i\cdot z^j)z_n^k\phi^-\rangle.
\label{cubicidentity}
\end{equation}

In order to obtain the standard of the Hamiltonian for 
the low-energy effective D-particle dynamics, 
it is sufficient to consider terms up to 4-th order in $(z, \bar{z})$, 
such as 
$
F[z, \bar{z}]=(\bar{z}^i\cdot z^j)(\bar{z}^i\cdot z^j). 
$
We find 
\[
\int [dU(N)] \, U(N)_{Na_1}U(N)_{Na_2}
U(N)^{-1}_{b_1N}U(N)^{-1}_{b_2N}(X^iX^j)_{a_1b_1}(X^iX^j)_{a_2b_2}
\]
\[
=(X^iX^j)_{a_1b_1}(X^iX^j)_{a_2b_2}
\]
\[
\times \Big(
{N\over N^2-1}(\delta_{NN}\delta_{NN}\delta_{a_1b_1}
\delta_{a_2b_2}+ \delta_{NN}\delta_{NN}\delta_{a_1b_2}
\delta_{a_2b_1})
\]
\[
-{1\over N^2-1}(
\delta_{NN}\delta_{NN}\delta_{a_1b_1}
\delta_{a_2b_2}
+\delta_{NN}\delta_{NN}\delta_{a_1b_2}
\delta_{a_2b_1})
\Big)
\]
\[
=(X^iX^j)_{a_1b_1}(X^iX^j)_{a_2b_2}\times {N-1\over N^2-1}
(\delta_{a_1b_1}\delta_{a_2b_2}+ \delta_{a_1b_2}\delta_{a_2b_1})
\]
\begin{equation}
=
{1\over N+1}\Big(
\Tr(X^iX^j)\Tr(X^iX^j)+\Tr(X^iX^jX^iX^j)
\Big). 
\end{equation}
Similarly, for 
$
F=(\bar{z}^i\cdot z^j)(\bar{z}^j\cdot z^i), 
$
we arrive at the expression, 
\begin{equation}
{1\over N+1}\Big(
\Tr(X^iX^j)\Tr(X^jX^i)+\Tr(X^iX^iX^jX^j)\Big). 
\end{equation}
These two examples of quartic $F$ enable us to represent the standard potential term 
\[
\Tr[X^i, X^j]^2 =2\Tr(X^iX^jX^iX^j-X^iX^iX^jX^j)
\]
 in terms of the D-particle field operators 
by
\begin{equation}
2(\langle \phi^+, \phi^-\rangle+1)
(\langle \phi^+, \Big((\bar{z}^i\cdot z^j)^2-(\bar{z}^i\cdot z^j)
(\bar{z}^j\cdot z^i) \Big)\phi^-\rangle ,
\end{equation}
which is 4-th order in the D-paricle field operators. 

Obviously,  any two bilinear operators in the case that there 
are no derivatives 
involved in $F$ are commutative as operations on 
the states, {\it i.e.}
\begin{equation}
[\langle \phi^+, F[z, \bar{z}]\phi^-\rangle , 
\langle \phi^+, G[z, \bar{z}]\phi^-\rangle ]
|N[X]\rangle 
=0, 
\end{equation}
even though the D-particle creation and annihilation operators 
do not satisfy the usual canonical commutation relations. 

\subsection{Bilinear operators with derivatives}
Let us next study cases in which $F$ involves derivatives. 
We first need some formulas describing the action of 
derivatives on $\delta$-functions, appearing when the 
annihilation operator $\phi^-$ acts on a general $N$-body 
state, as defined by \eqref{Nannihilation3}:
\[
\hspace{-5cm} {\partial\over \partial z_n^{(N)}} 
\delta^{2}\Big(z_n^{(N)}-(U(N)XU(N)^{-1})_{nN}\Big)\]
\begin{equation}=-U(N)_{Nb}\Big[{\partial\over \partial X_{ab}}
\delta^2\Big(z_n^{(N)}-(U(N)XU(N)^{-1})_{nN}\Big)\Big]
U(N)^{-1}_{an}, 
\end{equation}
\[
\hspace{-5cm} {\partial\over \partial \bar{z}_n^{(N)}} 
\delta^2\Big(z_n^{(N)}-(U(N)XU(N)^{-1})_{nN}\Big)
\]
\begin{equation}
=-U(N)_{nb}\Big[{\partial\over \partial X_{ab}}
\delta^2\Big(z_n^{(N)}-(U(N)XU(N)^{-1})_{nN}\Big)\Big]
U(N)^{-1}_{aN}
\end{equation}
with $n< N$. For $n=N$, we have 
\[
\hspace{-5cm} {\partial\over \partial x_N^{(N)}} 
\delta\Big(x_N^{(N)}-(U(N)XU(N)^{-1})_{NN}\Big)
\]
\begin{equation}
=-U(N)_{Nb}\Big[{\partial\over \partial X_{ab}}
\delta\Big(x_N^{(N)}-(U(N)XU(N)^{-1})_{NN}\Big)\Big]
U(N)^{-1}_{aN}.
\end{equation}

Let us consider the simplest case $F_n={\partial \over \partial x_n^i}$. When 
the corresponding bilinear operator
 acts on a general $N$-body state of the form $|\Psi\rangle 
=\int \frac{[d^dX]}{[dV(N)]}
\Psi[X]|N[X]\rangle$, we find 
\[
\langle \phi^+, {\partial \over \partial x_n^i}\phi^-\rangle|\Psi\rangle
 =\int \frac{[d^dX]}{[dU(N)]}\Psi[X]\int [D^2z]\phi^{+}[z, \bar{z}]
\frac{1}{2}(\partial_{z_n} +
\partial_{\bar{z}_n})
\]
\[\times
\int [dU(N)]
 \, \delta(x_N-X^{U(N)}_{NN})\delta^2(z_1-X^{U(N)}_{1N})
\delta^2(z_2-X^{U(N)}_{2N})
\cdots \delta^2(z_{N-1}-X^{U(N)}_{N-1, N})
\sigma_N
 \]
 \[
 \times {\cal P}_{N-1}\llbracket 
 \phi^+[X^{U(N)}_{1, N-1}, X^{U(N)}_{2, N-1}, \ldots, 
\bar{X}^{U(N)}_{1,N-1}, \bar{X}^{U(N)}_{2, N-1}, 
\ldots, X_{N-1, N-1}^{U(N)}]
\cdots \phi^+[X_{11}^{U(N)}] \rrbracket |0\rangle.
 \]
 Performing partial integrations over both $(z, \bar{z})$ and 
 the matrix variables $(X_{aN},X_{Na})$ with the help of the above 
 formulas, we see that this is not vanishing only when 
 $n=N$, owing to the group integration $[dU(N)]$. 
 Note that for $n>N$, it vanishes, due to the 
 projection condition for the creation operator. 
 The final result is 
 \begin{equation}
\langle \phi^+, {\partial \over \partial x_n^i}\phi^-\rangle|\Psi\rangle
 = \delta_{nN}\int \frac{[d^dX]}{[dV(N)]} \, 
 \Tr\Big({\partial \over \partial X^i}\Psi[X]\Big)|N[X]\rangle ,
 \label{singleder1}
\end{equation}  
which leads to the general $N$-independent expression
\begin{equation}
\langle \phi^+, \sum_{n=1}^{\infty}
{\partial \over \partial x_n^i}\phi^-\rangle|\Psi\rangle
 = \int \frac{[d^dX]}{[dV(N)]} \, 
 \Tr\Big({\partial \over \partial X^i}\Psi[X]\Big)|N[X]\rangle .
 \label{singleder2}
\end{equation}
This shows that the operator $\langle \phi^+, \sum_{n=1}^{\infty}
{\partial \over \partial x_n^i}\phi^-\rangle|\Psi\rangle$ 
is the generator corresponding to the global translation 
of D particles. It is interesting that in spite of the 
infinite sum $\sum_{n=1}^{\infty}
{\partial \over \partial x_n^i}$, it generates translations of only the diagonal 
components of the matrix coordinates $X^i$ for arbitrary $N$. 
Thus, we have the commuation relation
\[
[\langle \phi^+, \sum_{n=1}^{\infty} z^i_n\phi^-\rangle, 
\sum_{n=1}^{\infty}
\langle \phi^+, {\partial \over \partial x^j_n}\phi^-\rangle ]|\Psi\rangle
\]
\begin{equation}=
\int \frac{[d^dX]}{[dV(N)]} \, 
[\Tr X^i, \Tr\Big({\partial \over \partial X^j}\Big)]\Psi[X]|N[X]\rangle
 =-\delta_{ij}
\langle \phi^+, \phi^-\rangle|\Psi\rangle . 
\end{equation}
This is almost identical to the form which we would obtain if 
the creation and annihilation fields satisfied the 
canonical commutation relation, {\it except} for 
the normalization. More precisely, we would get the form 
above, except for an 
infinite constant multiplying the right-hand side. 
This potential difficulty that we would encounter when 
the ordinary canonical structure is adopted for our 
D-parrticle fields is avoided when we employ the projection conditions 
\eqref{projectioncreation} and \eqref{projectionannihilation}. 

Apart from this difference of the normalization constant, 
the operator  $\langle \phi^+, \sum_{n=1}^{\infty}
{\partial \over \partial x_n^i}\phi^-\rangle$ essentially 
plays the role of the translation generator, 
almost as in the 
case of a canonical field theory. To see this, it is useful to  
study its commutator with $
\langle \phi^+, \Big((\bar{z}^i\cdot z^j)^2-(\bar{z}^i\cdot z^j)
(\bar{z}^j\cdot z^i) \Big)\phi^-\rangle$, which corresponds to 
$\Tr([X^i, X^j]^2)/2$. 
If it is reduced to the matrix form, the commutator clearly 
vanishes. Hence, we must have 
\begin{equation}
[\langle \phi^+, \sum_{n=1}^{\infty}
{\partial \over \partial x_n^i}\phi^-\rangle
, \langle \phi^+, \Big((\bar{z}^i\cdot z^j)^2-(\bar{z}^i\cdot z^j)
(\bar{z}^j\cdot z^i) \Big)\phi^-\rangle]|\Psi\rangle=0. 
\end{equation}
On the other hand, we can attempt to carry out a 
global shift of the vector coordinates $z_n^i \rightarrow z_n^i +c^i$ directly in terms of the bilinear forms.  This 
transforms the function $F$ in $\langle \phi^+, \Big((\bar{z}^i\cdot z^j)^2-(\bar{z}^i\cdot z^j)
(\bar{z}^j\cdot z^i) \Big)\phi^-\rangle$ as 
\[F=
(\bar{z}^i\cdot z^j)^2-(\bar{z}^i\cdot z^j)
(\bar{z}^j\cdot z^i)
\]
\[
\rightarrow \delta F=
c^j\sum_{n=1}^{\infty} \Big(2(\bar{z}^j\cdot z^i)z^i_n+2(\bar{z}^i\cdot z^j)
\bar{z}^i_n) -
2(\bar{z}^i\cdot z^j)z^i_n
-2(\bar{z}^j\cdot z^i) \bar{z}^i_n)
\Big).
\]
Recalling the identity \eqref{cubicidentity}, we conclude that 
the bilinear form with this $\delta F$ vanishes. 
We can thus treat the bilinear form 
$\langle \phi^+, \sum_{n=1}^{\infty}
{\partial \over \partial x_n^i}\phi^-\rangle$ {\it almost} as if it were the generator of the global translation $z_n^i \rightarrow z_n^i + c^i$ for physical observables. 
In other words, the translation invariance of Yang-Mills theory 
is expressed in the second-quantized field theory 
formulation as symmetry under $z_n^i \rightarrow z_n^i + c^i$. 

As the next example for the operator $F$ involving derivatives, let us 
consider the second derivative 
\[
F_n={\partial \over \partial z_n^{i}}
{\partial \over \partial \bar{z}_n^{i}} , 
\] 
where only the sum over the spatial index $i$ is taken with fixed 
$n \, \,(<N)$. When the corresponding bilinear form 
acts on the $N$-body state $|\Psi\rangle$, we obtain
\[
\int \frac{[d^dX]}{[dU(N)]}\Psi[X]\int [dU(N)] \, 
U(N)_{Nb}U(N)_{an}^{-1}U(N)_{nb'}U(N)_{a'N}^{-1}\]
\[\times 
\delta(x_N-
X^{U(N)}_{NN}) )\, \delta^2(z_1-X^{U(N)}_{1N})\cdots 
\Big({\partial \over \partial X^i_{ab}}\delta(z_n-X^{U(N)}_{nN})\Big)
\Big({\partial \over \partial X^i_{a'b'}} \delta (\bar{z}_n-X^{U(N)}_{Nn})
\Big)
\]
\[ \times 
\cdots  \times \delta^2(z_{N-1}-X^{U(N)}_{N-1, N}) 
\phi^+[z, \bar{z}]\rhdllbra\phi^+[X_{N-1}^{U(N)}]\cdots \phi^+[X_1^{U(N)}]
\rrbracket |0\rangle
\]
\[
=\int \frac{[d^{d}X]}{[dU(N)]}\, 
\Big(
{\partial \over \partial X_{ab}^i}{\partial \over \partial X_{a'b'}^i}
\Psi[X]
\Big)
\Big(
{N\over N^2-1}\delta_{ab'}\delta_{ba'}-{1\over N^2-1}\delta_{a'b'}\delta_{ab}
\Big)|N[X]\rangle 
\]
\begin{equation}=
\int \frac{[d^dX]}{[dU(N)]}\Big[\frac{N}{N^2-1}\Tr\Big({\partial \over \partial X^i}{\partial \over \partial X^i}\Psi[X]\Big)
-\frac{1}{N^2-1}\Tr\Big({\partial \over \partial X^i}\Big)\Tr\Big({\partial \over \partial X^i}
\Big)\Psi[X]\Big]|N[X]\rangle.
\end{equation}
Note that this result is independent of $n$, as long as $n<N$. 
Similarly, for 
$
F_N=\Big({\partial \over \partial x_N^{i}}\Big)^2$, 
we find 
\[
\int \frac{[d^dX]}{[dU(N)]}\Psi[X]\int [dU(N)] \, 
U(N)_{Nb}U(N)_{aN}^{-1}U(N)_{Nb'}U(N)_{a'N}^{-1}
\]
\[\times 
\Big({\partial \over \partial X^i_{ab}}{\partial \over \partial X^i_{a'b'}}
\delta(x_N-
X^{U(N)}_{NN}) )\Big)
\delta^2(z_1-X^{U(N)}_{1N})\cdots 
\delta^2(z_n-X^{U(N)}_{nN}) 
\]
\[
\times 
\cdots  \times \delta^2(z_{N-1}-X^{U(N)}_{N-1, N}) 
\phi^+[z, \bar{z}]\rhdllbra\phi^+[X_{N-1}^{U(N)}]\cdots \phi^+[X_1^{U(N)}]
\rrbracket |0\rangle
\]
\[
=\int \frac{[d^{d}X]}{[dU(N)]}
\Big(
{\partial \over \partial X_{ab}^i}{\partial \over \partial X_{a'b'}^i}
\Psi[X]
\Big){N-1\over N^2-1}(\delta_{ba}\delta_{b'a'}+ \delta_{ba'}\delta_{b'a})
|N[X]\rangle
\]
\begin{equation}
=\frac{1}{N+1}
\int \frac{[d^dX]}{[dU(N)]}\Big[\Tr\Big({\partial \over \partial X^i}{\partial \over \partial X^i}\Psi[X]\Big)+\Tr\Big({\partial \over \partial X^i}\Big)\Tr\Big({\partial \over \partial X^i}
\Big)\Psi[X]\Big]|N[X]\rangle. 
\end{equation}
In order to obtain a universal form, it is appropriate to consider the 
bilinear form with the U($\infty$) Laplacian, 
$
4{\partial \over \partial \bar{z}^i }\cdot 
{\partial \over \partial z^i}, 
 $
 which reduces when acting on an $N$-body state to 
\[
4{\partial \over \partial \bar{z}^{i, (N)}}\cdot 
{\partial \over \partial z^{i, (N)}}
\rightarrow 4\sum_{n=1}^{N-1}
{\partial \over \partial z_n^{i, (N)}}
{\partial \over \partial \bar{z}_n^{i, (N)}}+
\Big({\partial \over \partial x_N^{i, (N)}}\Big)^2, 
\]
and hence we obtain
\[
\langle \phi^+, 4{\partial \over \partial \bar{z}^i }\cdot 
{\partial \over \partial z^i}\phi^-\rangle |\Psi\rangle 
={1\over N+1}\int \frac{[d^{d}X]}{[dU(N)]}
\Big[
(4N+1)\Tr\Big({\partial \over \partial X^i}{\partial \over \partial X^i}\Big)
\Psi[X]
\]
\begin{equation}
-3\, \Tr\Big({\partial \over \partial X^i}\Big)\Tr\Big({\partial \over \partial X^i}
\Big)\Psi[X]\Big]|N[X]\rangle .
\end{equation}
We can  rewrite this as
\[
(4N+1)\int \frac{[d^{d}X]}{[dU(N)]}\Big[\Tr\Big(
{\partial \over \partial X_{ab}^i}{\partial \over \partial X_{a'b'}^i}\Big)
\Psi[X]\Big]|N[X]\rangle 
\]
\[
=(N+1)\langle \phi^+, 4{\partial \over \partial \bar{z}^i }\cdot 
{\partial \over \partial z^i}\phi^-\rangle |\Psi\rangle
+3\int \frac{d^{d}X]}{[dU(N)]}\, \Big[\Tr\Big({\partial \over \partial X^i}\Big)
\Tr\Big({\partial \over \partial X^i}\Big)\Psi[X]\Big]
|N[X]\rangle.
\]
We also have 
\begin{equation}
\langle \phi^+, {\partial \over \partial \bar{z}^i}\phi^-\rangle 
\langle \phi^+, {\partial \over \partial z^i}\phi^-\rangle|\Psi\rangle
={1\over 4}\int \frac{d^dX]}{[dU(N)]}\, \Big[\Tr\Big({\partial \over \partial X^i}\Big)
\Tr\Big({\partial \over \partial X^i}\Big)\Psi[X]\Big]
|N[X]\rangle.
\end{equation}
This relation holds because, for the $N$-body states,  
only the term $\langle \phi^+, \partial_{x_n}\phi^-\rangle^2/4$ contributes, as seen from  \eqref{singleder1} and \eqref{singleder2}. 
 
Collecting all these results, we have derived the identity
\[
(4\langle \phi^+, \phi^-\rangle +1)\int \frac{[d^dX]}{[dU(N)]}
\Big[\Tr\Big(
{\partial \over \partial X^i}{\partial \over \partial X^i}\Big)
\Psi[X]
\Big]|N[X]\rangle \]
\begin{equation}
=4\Big[
(\langle\phi^+, \phi^-\rangle +1)\langle \phi^+, {\partial \over \partial \bar{z}^i }\cdot 
{\partial \over \partial z^i}\phi^-\rangle
+3\langle \phi^+, {\partial \over \partial \bar{z}^i}\phi^-\rangle 
\langle \phi^+, {\partial \over \partial z^i}\phi^-\rangle|\Big]\Psi\rangle.
\end{equation}
We also note that in our case of  a flat spacetime, we can assume that 
the total center-of-mass momentum vanishes, and hence we can set 
\begin{equation}
\Tr\Big({\partial \over \partial X^i}\Big)\Psi[X]=0, 
\label{cmass}
\end{equation}
or equivalently, 
\begin{equation}
\langle \phi^+, \partial_{z^i} \phi^-\rangle =0, 
\end{equation}
in order to simplify the formulas. 

\subsection{Schr\"{o}dinger equation}
We can now rewrite the Schr\"{o}dinger equation 
in terms of these bilinear operators. In configuration space, we have 
\[
i{\partial \over \partial t}\Psi[X]
=-\Tr\Big({g_s\ell_s\over 2}{\partial \over \partial X^i}
{\partial \over \partial X^i} +{1\over 4g_s\ell_s^5}
[X^i, X^j]^2
\Big)\Psi[X]
\]
\begin{equation}
=
-{1\over 2}\Tr\Big[g_s\ell_s{\partial \over \partial X^i}
{\partial \over \partial X^i} +{1\over g_s\ell_s^5}
(X^iX^jX^iX^j-X^iX^iX^jX^j)
\Big]\Psi[X].
\end{equation}
Here  we have suppressed the fermionic part, which is 
treated in the next section. 
In the second-quantized form, this is expressed as 
\begin{equation}
{\cal H}|\Psi\rangle =0, 
\end{equation}
\[
{\cal H}=i(4\langle \phi^+, \phi^-\rangle +1)\partial_t+
\]
\[
2g_s\ell_s\Big(
(\langle \phi^+, \phi^-\rangle +1) \langle \phi^+, 
\partial_{\bar{z}^i}\cdot \partial_{z^i}
\phi^-\rangle 
+3\langle \phi^+, \partial_{\bar{z}^i}\phi^-\rangle\cdot
\langle \phi^+, \partial_{z^i}\phi^-\rangle
\Big)
\]
\begin{equation}
+\frac{1}{2g_s\ell_s^5}
(4\langle \phi^+, \phi^-\rangle +1)(\langle \phi^+, \phi^-\rangle 
+1)\langle \phi^+, \Big(
(\bar{z}^i\cdot z^j)^2 -(\bar{z}^i\cdot z^j)(\bar{z}^j\cdot z^i)
\Big)\phi^-\rangle.
\end{equation}
In the large $N$ limit and in the center-of-mass frame, this is simplified to 
\[
i\partial_t|\Psi\rangle =\Big[
-{g_s\ell_s\over 2}\langle \phi^+, 
\partial_{\bar{z}^i}\cdot \partial_{z^i}
\phi^-\rangle 
-\frac{\langle \phi^+, \phi^-\rangle}{2g_s\ell_s^5}\langle \phi^+, \Big(
(\bar{z}^i\cdot z^j)^2 -(\bar{z}^i\cdot z^j)(\bar{z}^j\cdot z^i)
\Big)\phi^-\rangle\Big]
|\Psi\rangle .
\]

Obviously,  the Schr\"{o}dinger equation  commutes with the number operator 
$\langle \phi^+, \phi^-\rangle$ and  with the translation 
operator $\langle \phi^+, \sum_{n=1}^{\infty}\frac{\partial}
{\partial x_n^i}\phi^-\rangle$, according to the above 
discussion of the global translation in our
 infinite-dimensional base space.  We expect that after appropriate 
`bosonization' the former would be interpreted as 
a consequence of the gauge symmetry associated with the 
RR gauge fields in the bulk. 
We also note the following.  
As emphasized in some previous works \cite{stud0} by the 
present author, Yang-Mills quantum mechanics 
embodies the characteristic scales of 
D-particle dynamics in the form of 
scaling invariance under 
\[
X^i \rightarrow \lambda X^i, \quad t\rightarrow \lambda^{-1}t, 
\quad g_s\rightarrow \lambda^3 g_s, 
\]
conforming to the space-time uncertainty relation 
$\Delta t \Delta X \gtrsim \ell_s^2$ \cite{stu} as a 
qualitative characterization of 
the non-locality of string theory.  In the present 
field-theory formulation, the same scaling symmetry 
is satisfied under
\[
(z^i, \bar{z}^i)  \rightarrow \lambda (z^i,\bar{z}^i) \quad t\rightarrow \lambda^{-1}t, 
\quad g_s\rightarrow \lambda^3 g_s, 
\]
provided that the field operators possess  scaling such that 
the combination $\sqrt{[d^{2d}z]}\phi^{\pm}[z,\bar{z}]$ 
has zero scaling dimension. 
It is very desirable to clarify the geometrical 
meanings of all these symmetries in terms of some intrinsic language 
which is suitable for our infinite-dimensional base space.

\section{Supercoordinates}
\setcounter{equation}{0}
In this section, we briefly describe the extension 
of our formalism to fermionic matrix variables. 
They can be  
taken into account by introducing infinite-dimensional 
supercoordinates for D-particle fields, through a straightforward 
extension of what we have presented so far. 
Let us first recall the fermionic Hamiltonian in the 
configuration space, 
\begin{equation}
H_{\rm f}=\frac{1}{2}\Tr\Big(\Theta\Gamma^i[X^i, \Theta]\Big),
\end{equation}
where the components of the spinor 
 matrices, $\Theta$,  are Hermitian 
(in the sense of Majorana-Weyl)  and Grassmannian matrix 
coordinates, and the  quantities $\Gamma^i$ are the SO(9) 
Dirac matrices which are real and symmetric. 
The canonical anti-commutation relations are 
expressed as 
\begin{equation}
\{\Theta^{\alpha}_{ab}, \Theta_{cd}^{\beta}\}
=\delta_{\alpha\beta}\delta_{ad}\delta_{bc}, 
\end{equation}
with $\alpha, \beta, \ldots$ being the $16$-component 
spinor indices and $a, b, \ldots$ being the U$(N)$ indices,  
as in the case of bosonic variables. 
For our purpose, it is necessary to decompose the 
spinor coordinates into canonical conjugate pairs 
of generalized coordinates and momenta. This decomposition 
cannot be done covariantly with respect to SO(9) symmetry. 
One common way of doing this is to 
use SO(7)$\times$U(1) notation \cite{so7} by decomposing 
the spinors in the eigenstates of $\Gamma^9$. 

Since for the restricted purpose of the present paper,  
making SO(7) manifest has no particular 
merit,\footnote{
Of course, it is straightforward to adapt the following 
treatment to the 
SO(7)$\times$U(1) convention. 
}
let us carry out the most naive decomposition, 
\begin{equation}
S^A_{ab} =\frac{1}{\sqrt{2}}(\Theta^{2A-1}_{ab}+
i\Theta^{2A}_{ab}), 
\quad
\Pi^{A}_{ab}=\frac{1}{\sqrt{2}}(\Theta^{2A-1}_{ab}
-i\Theta^{2A}_{ab}), 
\end{equation}
with $A=1, 2, \ldots 8$. 
Then, we have
\begin{equation}
\{S^A, S^B\}=0=\{\Pi^{A},\Pi^{B}\}, \quad 
\{S^A, \Pi^{B}\}=\delta_{AB}, 
\end{equation}
suppressing the U($N$) indices. 
Thus, in terms of the 16 component spinor notation, we have
\begin{equation}
\begin{pmatrix}
S^1 \\
\Pi^1 \\
S^2 \\
\Pi^2 \\
. \\
. \\
. \\
S^8 \\
\Pi^8 \\
\end{pmatrix}
 =\frac{1}{\sqrt{2}}
\begin{pmatrix}
1 & i & 0 & 0 & 0 & . & . & 0 & 0 \\
1 & -i & 0 & 0 & 0 & . & . & 0 & 0 \\
0 & 0 & 1 & i & 0 & .  & . & 0 & 0 \\
0 & 0 & 1 & -i & 0 & . & . & 0 & 0 \\
. & . & . & . & . & . & . & . & . \\
. & . & . & . & . & . & . & . & . \\
0 & 0 & 0 & 0 & 0 & . & . & 1 & i \\
0 & 0 & 0 & 0 & 0 & . & . & 1 & -i \\
\end{pmatrix}\Theta\equiv L\Theta.
\end{equation}
Note that we have the following:
\begin{equation}
\Omega \equiv LL^t=\begin{pmatrix}
0 & 1 & 0 & 0 & 0 & . & . & 0 & 0 \\
1 & 0 & 0 & 0 & 0 & . & . & 0 & 0 \\
0 & 0 & 0 & 1 & 0 & .  & . & 0 & 0 \\
0 & 0 & 1 & 0 & 0 & . & . & 0 & 0 \\
. & . & . & . & . & . & . & . & . \\
. & . & . & . & . & . & . & . & . \\
0 & 0 & 0 & 0 & 0 & . & . & 0 & 1 \\
0 & 0 & 0 & 0 & 0 & . & . & 1 & 0 \\
\end{pmatrix},
\end{equation}
\begin{equation}
L^{-1}=\begin{pmatrix}
1 & 1 & 0 & 0 & 0 & . & . & 0 & 0 \\
-i & i & 0 & 0 & 0 & . & . & 0 & 0 \\
0 & 0 & 1 & 1 & 0 & .  & . & 0 & 0 \\
0 & 0 & -i & i & 0 & . & . & 0 & 0 \\
. & . & . & . & . & . & . & . & . \\
. & . & . & . & . & . & . & . & . \\
0 & 0 & 0 & 0 & 0 & . & . & 1 & 1 \\
0 & 0 & 0 & 0 & 0 & . & . & -i & i \\
\end{pmatrix}.
\end{equation}
Next, we define
\begin{equation}
\tilde{\Gamma}^i\equiv (L^t)^{-1}\Gamma^i L^{-1}=
(\tilde{\Gamma}^i)^t
\end{equation}
and 
\begin{equation}
\begin{pmatrix}
S^1 \\
\Pi^1 \\
S^2 \\
\Pi^2 \\
. \\
. \\
. \\
S^8 \\
\Pi^8 \\
\end{pmatrix}=
\begin{pmatrix}
S^1 \\
0 \\
S^2 \\
0 \\
. \\
. \\
. \\
S^8 \\
0 \\
\end{pmatrix}
+ 
\begin{pmatrix}
0 \\
\Pi^1 \\
0 \\
\Pi^2 \\
. \\
. \\
. \\
0 \\
\Pi^8 \\
\end{pmatrix}
\equiv 
\tilde{S} + \tilde{\Pi}.
\end{equation}
In terms of these variables, the fermionic Hamiltonian is 
\begin{equation}
H_{\rm f}=\frac{1}{2}\Tr\Big(
(\tilde{S}+\tilde{\Pi})\tilde{\Gamma}^i[X^i, \tilde{S}+\tilde{\Pi}]\Big)=
\frac{1}{2}\Tr\Big(
\tilde{S}\tilde{\Gamma}^i[X^i, \tilde{S}] +
\tilde{\Pi}\tilde{\Gamma}^i[X^i, \tilde{\Pi}]
+2\tilde{\Pi}\tilde{\Gamma}^i[X^i, \tilde{S}]\Big).
\label{fermihamilton}
\end{equation}

As in the case of bosonic coordinates, the components 
of the spinor variables which are relevant for $N$-th 
creation field operator are 
\begin{equation}
S_{aN}, \quad S_{Na}, \quad \Pi_{aN}=S_{aN}^{\dagger}, 
\quad 
\Pi_{Na}=S_{Na}^{\dagger}
\end{equation}
with $a\in (1, 2, \ldots, N)$. 
Unlike in the bosonic case, $(S_{aN}, S_{Na})$ 
and $(\Pi_{Na}, \Pi_{aN})$ for $a\ne N$ are 
two independent sets of complex spinor coordinates and their 
canonical momenta. 
We adopt a representation in which the 
$S$-variables are diagonalized. We then embed them 
into two sets of infinite-dimensional {\it complex} spinor vector 
coordinates $\zeta$ and  $\bar{\zeta}$,  such that the projection 
onto $N$-body states is done according to the relation 
\begin{equation}
\zeta^{(N)}_a=S_{aN}, \quad 
\bar{\zeta}^{(N)}_a=S_{Na} 
\end{equation}
with $a<N$, and with the projection conditions
\begin{equation}
\zeta_{N}^{(N)}=\bar{\zeta}_{N}^{(N)}   \quad \mbox{and} 
\quad 
\quad \zeta^{(N)}_{a}
=\bar{\zeta}^{(N)}_a=0 \quad \mbox{for} \quad a > N
\end{equation}
for the $(N)$-th creation superfield operator.  
For the annihilation superfield, the projection is 
made in momentum space according to 
\begin{equation}
\frac{\partial}{\partial {\zeta}_{N}^{(N)}}=\frac{\partial}{\partial \bar{\zeta}_{N}^{(N)} }  \quad \mbox{and} 
\quad 
\quad \frac{\partial}{\partial \zeta^{(N)}_{a}}
=\frac{\partial}{\partial \bar{\zeta}^{(N)}_a}
=0 \quad \mbox{for} \quad a > N.
\end{equation}
Keep in mind, again, that the spinor coordinates $\zeta$ and  $\bar{\zeta}$ are independent 
and {\it not}  complex conjugates of
 each other, in contrast to the bosonic 
coordinates.  The extensions of the projection conditions 
\eqref{projectioncreation} and \eqref{projectionannihilation} 
are, respectively, 
\begin{equation}
\phi^+[z, \bar{z}, \zeta, \bar{\zeta}]{\cal P}_n=
{\cal P}_{n+1}
\phi^+[\tilde{P}_nz, \tilde{P}_n\bar{z}, \tilde{P}_n\zeta, 
\tilde{P}_n\bar{\zeta}], 
\end{equation}
\begin{equation}
{\cal P}_n\phi^-[P_nz, P_n\bar{z}, 
P_n\zeta, P_n\bar{\zeta}]=\phi^-[z, \bar{z}, \zeta, \bar{\zeta}]{\cal P}_{n+1}, 
\end{equation}
using an obvious generalization of the projection operators $P_n$ 
and  $\tilde{P}_n$.

With the above conventions, we can now introduce the 
D-particle superfield
\[
\phi^{\pm}[z,\bar{z}] \rightarrow 
\phi^{\pm}[z, \bar{z}, \zeta, \bar{\zeta}]. 
\]
The $N$-body basis state and its gauge-statistics condition are 
\[
|N[X,S]\rangle 
={\cal P}_{N+1}\llbracket\phi^+[X_{1, N}, X_{2, N}, \ldots, X_{N-1,N}
\bar{X}_{1,N}, \bar{X}_{2, N}, 
\ldots,X_{N,N-1}, X_{N, N}, 
\]
\[
S_{1, N}, S_{2, N}, \ldots, S_{N-1,N},
S_{N,1}, S_{N,2}, 
\ldots, S_{N, N-1}, S_{N, N}]
\cdots \phi^+[S_{11}]\rrbracket |0\rangle 
\]
\[
={\cal P}_{N+1}\llbracket \phi^+[X_{1, N}^{U(N)}, X_{2, N}^{U(N)}, \ldots, 
\bar{X}_{1,N}^{U(N)}, \bar{X}_{2, N}^{U(N)}, 
\ldots, X_{N, N}^{U(N)}, 
\]
\begin{equation}
S_{1, N}^{U(N)}, S_{2, N}^{U(N)}, 
\ldots, S_{N-1,N}^{U(N)},
S_{N,1}^{U(N)}, S_{N,2}^{U(N)}, 
\ldots, S_{N, N}^{U(N)}]
\cdots \phi^+[S_{11}^{U(N)}]\rrbracket |0\rangle .
\end{equation}
The action of the annihilation superfield is 
\[
\phi^-[z,\overline{z},\zeta, \bar{\zeta}]{\cal P}_{N+1}\llbracket
 \phi^+[X_{1, N}, X_{2, N}, \ldots, X_{N-1,N}
\bar{X}_{1,N}, \bar{X}_{2, N}, 
\ldots,X_{N,N-1}, X_{N, N}, 
\]
\[
S_{1, N}, S_{2, N}, \ldots, S_{N-1,N},
S_{N,1}, S_{N,2}, 
\ldots, S_{N, N-1}, S_{N, N}]
\cdots \phi^+[S_{11}]\rrbracket |0\rangle 
\]
\[=
\int [dU(N)]\, \delta(x_N-
X^{U(N)}_{NN})\delta^2(z_1-X^{U(N)}_{1N})
\delta^2(z_2-X^{U(N)}_{2N})
\cdots \delta^2(z_{N-1}-X^{U(N)}_{N-1, N})
\]
\[
\delta(\zeta_N-
S^{U(N)}_{NN})\delta^2(\zeta_1-S^{U(N)}_{1N})
\delta^2(\zeta_2-S^{U(N)}_{2N})
\cdots \delta^2(\zeta_{N-1}-S^{U(N)}_{N-1, N})
\sigma_N
\]
\[\times {\cal P}_{N}\llbracket 
\phi^+[X_{1, N-1}^{U(N)}, X_{2, N-1}^{U(N)}, \ldots, 
\bar{X}_{1,N-1}^{U(N)}, \bar{X}_{2, N-1}^{U(N)}, 
\ldots, X_{N-1, N-1}^{U(N)}, 
\]
\begin{equation}
S_{1, N-1}^{U(N)}, S_{2, N-1}^{U(N)}, 
\ldots, S_{N-2,N-1}^{U(N)},
S_{N-1,1}^{U(N)}, S_{N-1,2}^{U(N)}, 
\ldots, S_{N-1, N-1}^{U(N)}]
\cdots \phi^+[S_{11}^{U(N)}]\rrbracket |0\rangle 
\end{equation}
where the two-dimensional 
$\delta$-function for Grassmann variables should be 
understood as
\begin{equation}
\delta^2(\zeta_n-S_{nN})
=\delta(\zeta_n-S_{nN})\delta(\bar{\zeta}_n-S_{Nn}), 
\end{equation}
and 
$\sigma_N$ represents the projection condition including the 
spinor variables: 
\begin{equation}
\sigma_N=\delta(y_N)\delta(\zeta_N-\bar{\zeta}_N)
\prod_{k>N}\delta^2(z_k)\delta^2(\zeta_k) .
\end{equation}
For a 1-body state, the basis state 
$\phi^+[z, \bar{z}, \zeta, 
\bar{\zeta}]|0\rangle={\cal P}_2\phi^+[X_{11}, S_{11}]|0\rangle$ 
has $2^8$ component fields, which form 
a $256\, \, (=128+128)$ dimensional 
Kaluza-Klein (in the 10-dimensions sense)
 supergravity-multiplet of D0-branes as usual. 
Construction of dual states is similar and is therefore omitted here. 

Construction of bilinear forms corresponding to invariants 
involving fermionic variables is a straightforward 
extension of the bosonic case. 
Let us first recall the $\delta$-function of Grassmann variables, 
\[
\delta(\zeta-\theta), 
\]
which is defined such that 
\[
\int d\zeta f(\zeta)\delta(\zeta-\theta)=\pm f(\theta)
\]
for an arbitrary Grassmann even or odd function $f(\zeta)=f_0+ f_1\zeta$, respectively. 
The explicit form is 
\[
\delta(\zeta-\theta)=\zeta-\theta.
\]
Thus, acting with a Grassmann derivative on the $\delta$ function, we have 
\begin{equation}
\frac{\partial}{\partial \zeta^{(N)}_n}\delta(\zeta^{(N)}_n
-(U(N)S U(N)^{-1})_{nN})=
-U(N)_{Nb}\frac{\partial}{\partial S_{ab}}\delta
(\zeta^{(N)}_n
-(U(N)S U(N)^{-1})_{nN})U(N)^{-1}_{an}, 
\end{equation}
just as in the bosonic case. This shows that 
the rule for constructing bilinear forms involving 
fermionic variables is essentially the same as in the 
bosonic case. 

For example, consider the fermionic Hamiltonian $H_{\rm f}$ 
\eqref{fermihamilton}. 
If we consider the third term, 
\[
\Tr\Big(\tilde{\Pi}\tilde{\Gamma}^i[X^i, \tilde{S}]\Big)
=\Tr\Big(\tilde{S}\tilde{\Gamma}^i[X^i, \tilde{\Pi}]\Big), 
\]
we should first study the bilinear form 
\begin{equation}
\langle \phi^+, F \phi^-\rangle =\int [d^2z d^2\zeta]\, 
\phi^+[z,\bar{z}, \zeta, \bar{\zeta}] \triangleright F 
\phi^-[z, \bar{z}, \zeta, \bar{\zeta}], 
\end{equation}
with 
\[
F=(\bar{z}^i\cdot \zeta-\bar{\zeta}\cdot z^i)\tilde{\Gamma}^i
\frac{\partial}{\partial \zeta_n}, 
\]
or (equivalently)
\[
F=(\bar{z}^i\cdot \zeta-\bar{\zeta}\cdot z^i)\tilde{\Gamma}^i
\frac{\partial}{\partial \bar{\zeta}_n}.
\]
Then, as the action on the wave function, 
this gives 
\[
\int [dU(N)] U(N)_{Na}(X^iS-S X^i)_{ab}
U(N)^{-1}_{bN}U(N)_{Nd}U(N)^{-1}_{cn}
\tilde{\Gamma}^i\frac{\partial}
{\partial S_{cd}}\Psi[X,S]
\]
\[
=\int [dU(N)] U(N)_{Na}U(N)_{Nd}
U(N)^{-1}_{bN}U(N)^{-1}_{cn}
(X^iS-S X^i)_{ab}
\tilde{\Gamma}^i \frac{\partial}
{\partial S_{cd}}\Psi[X,S]
\]
\[
=\Big(\frac{N}{N^2-1}
(\delta_{NN}\delta_{Nn}\delta_{ab}\delta_{dc}
+\delta_{Nn}\delta_{NN}\delta_{ac}\delta_{db})
-\frac{1}{N^2-1}
(\delta_{Nn}\delta_{NN}\delta_{ab}\delta_{dc}
+\delta_{NN}\delta_{Nn}\delta_{ac}\delta_{db})\Big)
\]
\[
\times 
(X^iS-S X^i)_{ab}
\tilde{\Gamma}^i\frac{\partial}
{\partial S_{cd}}\Psi[X,S]
\]
\[=\delta_{Nn}
\frac{1}{N+1}\Tr\Big(
[X^i, S]
\tilde{\Gamma}^i{\partial\over \partial S}
\Big)\Psi[X, S], 
\]
with 
\[
\Big(\frac{\partial}{\partial S}\Big)_{ba}
=\frac{\partial}{\partial S_{ab}}=\Pi_{ba}.
\]
This leads to 
\[
(\langle \phi^+, \phi^-\rangle +1) \langle \phi^+, \sum_{n=1}^{\infty}(\bar{z}^i\cdot \zeta-\bar{\zeta}\cdot z^i)\tilde{\Gamma}^i
\frac{\partial}{\partial \zeta_n}\phi^-\rangle |\Psi\rangle 
\]
\[
=\int \frac{[d^9Xd^{16}S]}{[dU(N)]} 
\Tr\Big(
[X^i, S]
\tilde{\Gamma}^i{\partial\over \partial S}
\Psi[X, S]\Big)|N[X, S]\rangle . 
\]
Similarly, we obtain the fermionic part of the Hamiltonian as 
\[
(\langle \phi^+, \phi^-\rangle +1) \sum_{n=1}^{\infty}\Big[\frac{1}{2}\langle \phi^+, 
\Big(
(\bar{z}^i\cdot \zeta-\bar{\zeta}\cdot z^i)\tilde{\Gamma}^i
\zeta_n
-(z^i\cdot \frac{\partial}{\partial\zeta}-\frac{\partial}{\partial \bar{\zeta}}\cdot \bar{z}^i)
\tilde{\Gamma}^i\frac{\partial}{\partial \zeta}_n\Big)\phi^-\rangle 
\]
\[
+\langle \phi^+, (\bar{z}^i\cdot \zeta-\bar{\zeta}\cdot z^i)\tilde{\Gamma}^i
\frac{\partial}{\partial \zeta_n}\phi^-\rangle\Big]|\Psi\rangle 
\]
\begin{equation}
=
\int \frac{[d^9Xd^{16}S]}{[dU(N)]} 
\Tr\Big[\Big(
\frac{1}{2}[X^i, S]\tilde{\Gamma}^i S
+ \frac{1}{2}[X^i, \frac{\partial}{\partial S}]\tilde{\Gamma}^i 
\frac{\partial}{\partial S}+ 
[X^i, S]
\tilde{\Gamma}^i{\partial\over \partial S}
\Big)\Psi[X, S]\Big]|N[X, S]\rangle .
\end{equation}
All these bilinear forms of the super D-particle fields 
are, of course,  A-class and 
satisfy the standard operator algebras of the 
corresponding gauge invariants. It is also a straightforward task 
to construct 
supersymmetry generators with bilinears of 
Grassmann odd operators $F$, although we do not 
elaborate further in the present paper. 

\section{Concluding remarks}
We have proposed a field theory 
for D-particles by second-quantizing  (super) Yang-Mills U($N$) 
quantum mechanics  in the Fock space containing all different $N$ simultaneously. The  fact that the configuration space 
of $N$-body D particles is described by matrix coordinates 
requires that we construct an entirely new framework 
of quantum D-particle fields whose algebraic structure cannot be 
described by the ordinary canonical formalism. 
The scope of the present paper is to present a general 
framework in order to 
rewrite the content of Yang-Mills quantum mechanics in terms 
of the D-particle fields as an initial step towards our goal.  
Various desirable extensions and possible applications, as well as 
further clarifications 
of the formalism, are left for future works. Among such 
potential studies, it is worthwhile to specifically mention
 explorations of the algebraic 
structure of the formalism and, especially, extensions to covariant 
formulations in both 10 and 11 dimensional senses. In particular, 
it is  quite challenging  to ask possible 
applications of our ideas to M-theory interpretations \cite{bfss}. 

One of the main motivations for the present project was to  reformulate   
the duality between 
open and closed strings, in an analogy to 
the well-known Mandelstam duality in two-dimensional 
field theories. The field theories of D-branes are conjectured to 
form a new element,  which,  together with standard 
open and closed string field theories, constitutes a {\it trinity} of dualities associated with the 
open-closed string duality,  as illustrated by the diagram in Fig. 2. 
What we have discussed corresponds, in this diagram, 
 to the arrow on the left 
connecting the Yang-Mills (or open string, if we could include 
all massive open string modes) 
formulation to D-brane field theories. \vspace{0.6cm}
\begin{figure}[htbp]
\begin{center}
\includegraphics[width=13cm]{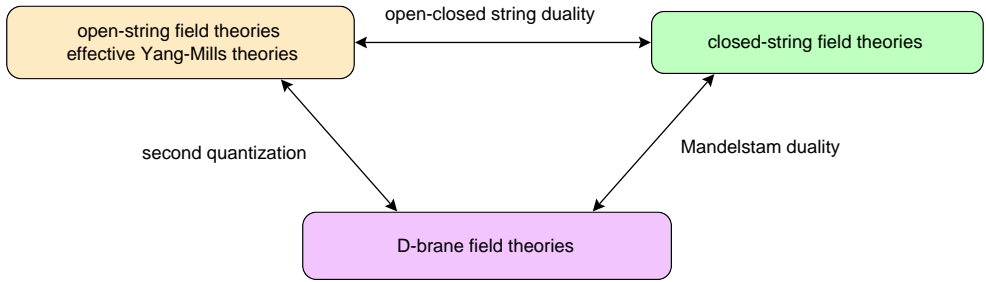}
\caption{\footnotesize{
A trinity of dualities: D-brane field theories point toward 
a third possible formulation of string theory treating 
D-branes as elementary excitations. 
}}
\end{center}
\vspace{-0.2cm}
\end{figure}

An expectation suggested here is that if we succeed in 
constructing an analog of the soliton operator for D particles 
in quantized closed string field theory, it would be described 
in the framework proposed in this paper.  Recently, an 
attempt to construct a soliton operator for D-branes 
has been reported \cite{ishibashi} within the framework of 
a special version (the so-called 
OSp-invariant string field theory) of 
closed string field theory. Unfortunately, however, it is difficult 
in the formulation of that work to see any signature of open-string degrees 
of freedom as collective coordinates of D-branes and the associated 
gauge structure.  An appropriate understanding 
of this latter aspect in many-body  D-brane systems
 should allow us to 
study the arrow on the right. An interesting problem 
is to determine what 
soliton operators look like in quantized closed string 
field theory of the type advocated in Ref. \cite{zwiebach}. 

Conversely, we may attempt to derive closed string field theories 
starting from our D-particle field theory by developing a `bosonization' technique 
for the D-brane fields. 
From this viewpoint, 
it would be interesting to investigate some suitable toy 
models as a warm-up exercise. Namely we may consider similar questions of a dual description 
for non-commutative solitons, where the classical solutions are
 known to exhibit a projector property, which is most probably 
 related to the structure we have discussed,  gives rise 
 \cite{noncommutative} to a Yang-Mills-like 
 symmetry structure. 
 In connection to this, it would also be interesting to 
study the BMN-type operators of the Yang-Mills theory 
of D-particles by using the present formalism. 
In our previous works, we have given some 
nontrivial  predictions \cite{sy, asy} for the correlation functions of 
the BMN-type operators from supergravity via holography, 
which may be useful to determine the correspondence between our 
D-particle field theory and closed string field theories.

 Finally, we touch on an interpretation 
 of the peculiar base space of the present D-particle 
 field theory. In terms of 
 string theory, the infinite number of vector 
 components $z^i_n$ and $\bar{z}^i_n$ correspond to 
 the coordinates of a D-particle and infinite degrees of freedom representing
  possible open strings emanating from it. In this sense, 
the base-space coordinates describe certain {\it fuzzy} 
(permeable and osmotic) spatial domains. 
These open string degrees of freedom are latent in their nature, since 
they are activated only when acting on states, depending on the number of D-particles. 
This novel feature was taken into account by the projection 
conditions. 
In a broad sense, the D-particle field theory 
is somewhat reminiscent of the idea of 
`elementary domains' \cite{yukawa} proposed by Yukawa 
in the late 1960s.  
Unlike the latter idea, where the non-locality  is governed by the explicit form of  extended  
domains, however, the non-locality in our theory caused by the open-string degrees of freedom is 
consistent with a more dynamical non-local structure of string theory as being characterized qualitatively 
by the space-time uncertainty relation \cite{stu}.  This is  responsible for the fact that D-particle Yang-Mills quantum mechanics encompasses 
general relativity, as we have demonstrated by deriving the 
3-body gravitational force among D-particles in Ref. \cite{3body}. 
It would be very nice if we could see a more direct link to 
supergravity in our field-theory language.

\vspace{0.5cm}
\noindent
Acknowledgements

A preliminary form of this paper was reported at the 21st 
Nishinomiya-Yukawa Memorial Symposium (YITP, Kyoto, November, 2006) and in the 
Komaba 2007 Workshop (Univ. of Tokyo, February, 2007). 
The author would like to thank the organizers and participants 
for their interest. 

The present work  is supported in part by Grants-in-Aid for Scientific Research [No. 13135205 (Priority Areas) and No. 16340067 (B)] from the Ministry of  Education, Science and Culture of Japan, and also by 
the Japan-US Bilateral Joint Research Projects  from JSPS.

\end{document}